\documentstyle[12pt]{article}

\newcommand{\be}{\begin{equation}}
\newcommand{\ee}{\end{equation}}
\newcommand{\bea}{\begin{eqnarray}}
\newcommand{\eea}{\end{eqnarray}}
\newcommand{\norsl}{\normalsize\sl}
\newcommand{\norsc}{\normalsize\sc}

\def \ksl {k \kern-.45em{/}}
\def \ppsl {p \kern-.45em{/}}
\def \nsl {n \kern-.45em{/}}

\def\theequation{\arabic{section}.\arabic{equation}}

\begin{document}


\begin{titlepage}

\title{The Gluon Self-Energy in the Coulomb \\
and Temporal Axial Gauges \\ 
via the Pinch Technique}

\author{
{\norsc Massimo Passera}\thanks{e-mail address: 
passera@mafalda.physics.nyu.edu}  {\norsl and} 
{\norsc Ken Sasaki}\thanks{Permanent address: 
Dept. of Physics, Yokohama National University, Yokohama 240, JAPAN.
e-mail address: sasaki@mafalda.physics.nyu.edu   
or sasaki@ed.ynu.ac.jp }\\
\norsl  Dept. of Physics, New York University\\
\norsl  4 Washington Place, New York, New York 10003, U.S.A.\\}

\date{}
\maketitle
 
\begin{abstract}
{\normalsize The $S$-matrix pinch technique is used to derive an effective 
gluon self-energy to one-loop order, when the theory is quantized 
in the Coulomb gauge (CG) and in the temporal axial gauge (TAG). 
When the pinch contributions are added, the gluon self-energies calculated 
in CG and TAG turn out to be identical and coincide with the result 
previously obtained with covariant gauges.
The issue of gauge independence of several quantities in hot QCD is 
discussed from the pinch technique point of view. It is also pointed out that 
the spurious singularities which appear in TAG calculations 
cancel out once the pinch contributions are combined.} 
\end{abstract}
 
\begin{picture}(5,2)(-290,-600)
\put(2.7,-115){NYU-TH-96/05/04}
\end{picture}

\thispagestyle{empty}
\end{titlepage}
\setcounter{page}{1}
\baselineskip 18pt

\setcounter{equation}{0}
\section{Introduction}
\smallskip

The $S$-matrix pinch technique (PT) is an algorithm which enables us to  
construct gauge-independent (GI) modified off-shell $n$-point functions 
through the rearrangement of Feynman graphs contributing to certain 
physical $S$-matrix elements. First introduced by 
Cornwall~\cite{rCa} some time ago to form the new GI-QCD proper 
vertices and propagators for the Schwinger-Dyson equations, 
the PT was used to obtain the one-loop 
GI effective gluon self-energy and vertices in 
QCD~\cite{rCP}\cite{rPapaQCD}. It has then been 
extensively applied to the standard model~\cite{StandardModel}.
Recently the PT was applied also to QCD at high temperature 
to calculate the gap equation for the magnetic mass~\cite{Nair} and 
to obtain the GI thermal $\beta$ function~\cite{Sasakia}\cite{Sasakib}. 

Indeed, the PT algorithm has scored a success in its applications 
to various fields. However, we can hardly say that it was fully 
understood and well established. In particular, since in the $S$-matrix PT
the effective amplitudes are obtained through the 
rearrangement of Feynman graphs, their uniqueness is at stake. 
One may argue that arbitrary pieces can always be moved around by hand from 
the vertex or box diagrams, as long as one does not alter the unique 
$S$-matrix element.  On the other hand, 
the $S$-matrix PT algorithm is expected to give rise to 
the same answers, even when one may choose an $S$-matrix element for 
a different process or start calculations with different gauge-fixing choices. 
Unfortunately, there exists so far no general proof on this point, and 
therefore, we may have to examine individual cases to convince ourselves 
of the validity of the PT algorithm. 
The process-independence of the PT has been recently 
proved~\cite{Watson} via explicit one-loop calculations. 
The independence of the gauge-fixing choices 
has been shown for the case of the effective gluon self-energy at 
one-loop order in the covariant gauge~\cite{rCP}, 
the background field gauge~\cite{Hiroshima}\cite{Papavass} 
and one of the non-covariant gauges, namely, the light-cone gauge~\cite{rCa}. 
However,  the PT calculations have not been carried out 
in the other interesting non-covariant gauges up to the present.

Non-covariant gauges such as the Coulomb gauge (CG) and the axial gauges
have long been used, both for theoretical analyses and for various 
numerical  calculations in  gauge theories~\cite{Noncov}.
These gauges are sometimes called ``physical'' gauges since in these gauges 
there is a close correspondence between independent fields and 
``physical'' degrees of freedom. In particular, the CG and 
the temporal axial gauge (TAG) have been often chosen for 
the perturbative calculations of QCD at finite 
temperature~\cite{Hot}-\cite{BP}. 
The reasons for these gauges being used are, for CG, that it is a natural 
gauge choice for the study of interactions between charges and, for TAG, that 
for a thermal system the rest frame of the heat bath singles out the 
four-vector $n^{\mu}=(1,0,0,0)$~\cite{Kajantie}. 

The gluon self-energy is a gauge-dependent quantity. 
Its one-loop expression in CG differs from the one in TAG. 
And the transversality relation is satisfied by the 
one-loop gluon self-energy calculated in TAG but not by the one 
in CG. However, the hard thermal loop $\delta \Pi_{\mu\nu}$ 
in the gluon self-energy is gauge independent, which means that 
$\delta \Pi_{\mu\nu}$'s calculated  in CG and in TAG are the same. 
The electric mass $m_{el}$, relevant for electric screening, and the 
``effective gluon mass'' $m_G$ in hot QCD are gauge 
independent quantities and they can be obtained from the 
one-loop gluon self-energy calculated in any gauge choice. 
Meanwhile it is well known that in TAG calculations there appear 
spurious singularities  which are  
due to the unphysical poles of   
$(k \cdot n)^ {-\lambda} , \lambda = 1,2$ in the TAG gluon 
propagator~\cite{Noncov}\cite{LeibbrandtA}. 
Several methods have been proposed to circumvent these singularities, 
and most noticeable are the 
principal-value prescription~\cite{PV}, 
the $n^{*}_{\mu}$-prescription~\cite{LeibbrandtB} 
and the $\alpha$-prescription~\cite{Landshoff}.

In this paper we apply the $S$-matrix PT and 
calculate an effective gluon self-energy to one-loop order 
in CG and TAG. The one-loop gluon self-energies both in CG and TAG have 
very complicated expressions. Even in these gauges  
we find that once the pinch contributions are added, 
we indeed obtain the same result for the effective gluon self-energy 
as the one derived before in different gauge choices. 
This gives another support for the usefulness of the $S$-matrix PT. 
We can also argue why the transversality relation holds for 
the gluon self-energy calculated in TAG, but not for the one in CG,
from the analysis of the structure of the pinch contributions. 
Moreover, we can explain why the thermal loops, the electric mass $m_{el}$ and the 
effective gluon mass $m_G$ in hot QCD are gauge independent from 
a simple inspection of the pinch contributions. 
Concerning the spurious singularities which appear in the gluon 
self-energy in TAG, we point out that these singularities also 
appear in the pinch contributions and they exactly cancel 
against the counterparts in the gluon self-energy. To show 
explicitly how these cancellations occur, we calculate  in TAG 
the one-loop gauge-independent thermal $\beta$ function 
$\beta_T$ in hot QCD.

The paper is organized as follows. In the next section, we develop 
the general prescription necessary for extracting the pinch contributions 
to the gluon self-energy from 
the one-loop quark-quark scattering amplitude. 
To establish our notation and to illustrate how to use the 
prescription developed in the previous section, we briefly 
review, in Sec. 3, the derivation of the 
pinch contribution to the gluon self-energy in the Feynman gauge (FG).
In Sec. 4 we calculate both the gluon self-energy and the pinch 
contribution in one-loop order in CG with an arbitrary gauge 
parameter $\xi_C$, and show that when combined they give the 
same expression for the effective gluon self-energy as the one 
obtained before in different gauge choices. In Sec. 5 
the similar calculations are performed in TAG with an arbitrary gauge 
parameter $\xi_A$. Also we calculate the thermal $\beta$ function $\beta_T$ 
at one-loop order in TAG and show how the spurious singularities 
appearing in the TAG gluon self-energy cancel against the counterparts in 
the pinch contribution. Sec. 6 is devoted to 
summary and discussion. In addition, we present three Appendices. 
In Appendix A we first give the one-loop pinch contributions 
to the gluon self-energy in CG with $\xi_C \ne 0$ 
from the vertex 
diagrams of the first and second kind and from box diagrams, separately. 
Then we give the expression of the pinch contribution rewritten in terms of 
different tensor bases. In Appendix B, we give the similar 
expressions calculated in TAG with $\xi_A \ne 0$. 
In Appendix C we list the formulae for thermal one-loop integrals 
necessary for calculating $\beta_T$ in TAG in Sec.5. 

\bigskip

\setcounter{equation}{0}
\section{Pinch Technique}
\smallskip

In this section we explain how to obtain the one-loop pinch 
contributions to 
the gluon self-energy. Let us consider the $S$-matrix element 
$T$ for the elastic quark-quark scattering at one-loop order in the 
Minkowski space, assuming that quarks have the same mass $m$. 
Throughout this paper we use the metric $(+,-,-,-)$.
Besides the self-energy diagram in Fig.1(a), 
the vertex diagrams of the first and second kind   
and the box diagrams contribute to $T$.
They are shown in Fig.2(a), Fig.3(a), and Fig.4(a), respectively.
These contributions are, in general, gauge-dependent, while the sum 
is gauge-independent. Then 
we single out the ``pinch parts'' of the vertex and box diagrams, 
which are depicted in Fig.2(b), Fig.3(b), and Fig.4(b). 
They emerge when a $\gamma^{\mu}$ matrix on the 
quark line is 
contracted with a four-momentum $k_{\mu}$ offered by a gluon propagator or a  
bare three-gluon vertex. Such a term triggers an elementary Ward 
identity of the form
\be
      \ksl = ({\ppsl} + \ksl -m) - ({\ppsl} -m).
\label{Ward}
\ee
The first term removes (pinches out) the internal quark propagator, 
whereas the second term vanishes on shell, or {\it vice versa} . This 
procedure leads to 
contributions to $T$ with one or two less quark propagators and, 
hence, we will call these 
contributions $T_P$,  ``pinch parts'' of $T$.

Next we extract from $T_P$ the pinch contributions to the gluon self-energy  
$\Pi^{\mu \nu}$. 
First note that the contribution of the gluon self-energy diagram to 
$T$ is written in the form (see Fig.1(a))
\be
  T^{(S.E)}=[T^a \gamma ^{\alpha}]D_{\alpha \mu}(k)
       \Pi^{\mu\nu}D_{\nu \beta}(k)[T^a \gamma ^{\beta}],
\ee
where $D(k)$ is a gluon propagator, $T^a$ is a representation matrix  
of $SU(N)$, and $\gamma ^{\alpha}$ and $\gamma ^{\beta}$ are 
$\gamma$ matrices on the external quark lines. 
The pinch contribution $\Pi_P^{\mu \nu}$ to 
$T_P$ should have the same form. 
Thus we must take away $[T^a \gamma ^{\alpha}]D_{\alpha \mu}(k)$ and 
$D^{\nu \beta}(k)[T^a \gamma ^{\beta}]$ from $T_P$.  
For that purpose we use the following identity satisfied by 
the gluon propagator and its inverse:
\bea
   g_{\alpha}^{\beta}&=&D_{\alpha \mu}(k)[D^{-1}]^{\mu \beta}(k)
           =D_{\alpha \mu}(k)[-k^2d^{\mu \beta}] + k_{\alpha}~  {\rm term} 
       \nonumber \\
           &=&D^{-1}_{\alpha \mu }(k)D^{\mu \beta}(k) =
             [-k^2 d_{\alpha \mu}]D^{\mu \beta}(k) + k_{\beta}~ {\rm term},
\label{Identity}
\eea
where
\be
     d^{\mu \nu} = g^{\mu \nu} - \frac{k^{\mu}k^{\nu}}{k^2}.  
\ee 
The $k_{\alpha}$ and  $k_{\beta}$ terms give  null results  when 
they are contracted with $\gamma_{\alpha}$ and of $\gamma_{\beta}$,
respectively,  of the external quark lines.

The pinch part of the one-loop vertex diagrams of the first kind depicted 
in Fig.2(b) plus their mirror graphs has a form 
\be
  T_P^{(V_1)}={\cal A}[T^a \gamma ^{\alpha}]D_{\alpha \beta}(k)
      [T^a \gamma ^{\beta}]~,
\ee
where ${\cal A}$ (also ${\cal B}_0$, ${\cal B}_{ij}$, ${\cal C}_0$, and 
${\cal C}_{ij}$ in the equations below) contains a loop integral. 
Using Eq.(\ref{Identity}) we find 
\be
    \gamma^{\alpha}D_{\alpha \beta}(k)\gamma ^{\beta}=
       \gamma^{\alpha}D_{\alpha \mu}(k)[-k^2 d^{\mu \nu}]
     D_{\nu \beta}(k)\gamma^{\beta}.
\label{gammagamma}
\ee
Thus  the contributions to  $\Pi^{\mu \nu}$ from the vertex diagrams of 
the first kind are written as 
\be
   \Pi^{\mu \nu (V_1)}_P=[-k^2 d^{\mu \nu}]{\cal A}.
\ee
 
The pinch part of the one-loop vertex diagrams of the second kind depicted 
in Fig.3(b) has a form 
\be
  T_P^{(V_2)}=\biggl[T^a \biggl\{ [\gamma^{\kappa}] {\cal B}_0 + 
     \sum_{i,j}{\cal B}_{ij} [\ppsl_i] p_j^{\kappa}  \biggr\} \biggr]
   D_{\kappa \beta}(k) [T^a \gamma ^{\beta}]~,
\label{TPV2}
\ee    
where $p_i$ and $p_j$ are four-momenta appearing 
in the diagrams. By redefinition of the loop-integral momentum
we can choose $p_i, p_j=p\  {\rm or}\  n$ in the cases of CG and TAG  
where $p$ is the 
loop-integral momentum and $n$ is a unit vector $n^{\mu}=(1,0,0,0)$ 
appearing in the CG and TAG gluon propagators. 
Using Eq.(\ref{gammagamma}) and 
\be
  [\ppsl_i] p_j^{\kappa}D_{\kappa \beta}(k)=
   [\gamma^{\alpha}]D_{\alpha \mu}(k)[-k^2 d^{\mu \lambda}]p_{i \lambda} 
    p_j^{\nu}D_{\nu \beta}(k)~,
\ee
we obtain  for the contributions to  $\Pi^{\mu \nu}$ from the vertex 
diagrams of the second kind
\bea
   \Pi^{\mu \nu (V_2)}_P&=&[-k^2 d^{\mu \nu}]{\cal B}_0 +
        [-k^2 d^{\mu \lambda}]\sum_{i,j}{\cal B}_{ij} p_{i \lambda} p_j^{\nu} 
       \nonumber \\
      & &+ (\mu \leftrightarrow  \nu)~,
\label{PIV2}
\eea
where $(\mu \leftrightarrow  \nu)$ terms are the contributions from 
mirror diagrams.
A further simplification can be made by using a formula
\be
    k^2 p_j^{\nu}=k^2 d^{\nu \tau}p_{j \tau}+k^{\nu}(kp_j).
\label{Formula}
\ee

The pinch part of the one-loop box diagrams depicted 
in Fig.4(b) has a form 
\be
  T_P^{(Box)}=[T^a] \biggl\{ [\gamma^{\alpha}] [\gamma_{\alpha}]{\cal C}_0 + 
     \sum_{i,j} {\cal C}_{ij} [\ppsl_i][\ppsl_j]  \biggr\} [T^a].
\ee
Again from Eq.(\ref{Identity}) we see that 
$[\gamma^{\alpha}] [\gamma_{\alpha}]$ and $[\ppsl_i][\ppsl_j]$ are rewritten 
as     
\be
  [\gamma^{\alpha}] [\gamma_{\alpha}] =
     [\gamma^{\alpha}]D_{\alpha \mu}(k)[k^4 d^{\mu \nu}] 
           D_{\nu \beta}(k)[\gamma_{\beta}]              
\ee
\be
    [\ppsl_i][\ppsl_j] =
        [\gamma^{\alpha}]D_{\alpha \mu}(k)[k^4 d^{\mu \lambda}
                d^{\nu \tau} p_{i \lambda} p_{j \tau}]
     D_{\nu \beta}(k)[\gamma_{\beta}]
\ee
and thus we obtain for the contributions to  $\Pi^{\mu \nu}$ from the 
box diagrams
\be
   \Pi^{\mu \nu (Box)}_P=[k^4 d^{\mu \nu}]{\cal C}_0 +
        [k^4 d^{\mu \lambda}d^{\nu \tau}]
        \sum_{i,j}{\cal C}_{ij} p_{i \lambda} p_{j \tau}~.
\ee

It is observed that the prescription developed here is general and can be 
applied to the calculation of the one-loop pinch contributions in any gauge.

\bigskip

\setcounter{equation}{0}
\section{PT Gluon Self-Energy in the Feynman Gauge}
\smallskip

In order to establish our notation, in this section we briefly review  
the derivation of the effective gluon self-energy  
in the Feynman gauge (FG) (the covariant gauge with $\xi = 1$). 
In the following we discuss the gluon self-energy both at 
$T=0$ and at 
finite temperature. In both cases we use the same notation $\int dp$ 
for the loop integral. At $T=0$ the loop integral should read as
\be
  \int dp = -i \mu^{4-D}\int \frac{d^Dp}{(2\pi)^D} 
,
\label{loop}
\ee
where $\mu$ is the 't Hooft mass scale,
while at finite temperature we use the imaginary time formalism 
of thermal field theory, and the loop integral should read as 
\be
    \int dp = \int \frac{d^3p}{8\pi^3} T \sum_n \qquad 
     ( {\rm imaginary\  time\  formalism}),
\ee
where the summation goes over the integer $n$ in $p_0 = 2\pi i n T$.

In FG the gluon propagator, 
$iD_{ab(FG)}^{\mu \nu}=i\delta _{ab}D_{(FG)}^{\mu \nu}$,  
has a very simple form
\be
   D_{(FG)}^{\mu \nu}(k)=\frac{-1}{k^2}
                 g^{\mu \nu},
\ee
and the three-gluon vertex is expressed as
\be
\Gamma^{abc}_{\lambda \mu \nu } (p, k, q) = -g f^{abc} 
              \biggl[ \Gamma^{P}_{\lambda \mu \nu}(p, k, q) 
              + \Gamma^{F}_{\lambda \mu \nu}(p, k, q) \biggr],
\label{VertexFG}
\ee
where 
\bea
     \Gamma^{P}_{\lambda \mu \nu}(p, k, q) &=& 
               p_{\lambda}g_{\mu \nu} -q_{\nu} g_{\lambda \mu} \nonumber \\
     \Gamma^{F}_{\lambda \mu \nu}(p, k, q) &=&
            2k_{\lambda}g_{\mu \nu}-2k_{\nu}g_{\lambda \mu} 
             -(2p+k)_{\mu}g_{\lambda \nu}, 
\label{Vertex}
\eea
and $f^{abc}$ are the structure constants of the group $SU(N)$. 
In the vertex each momentum flows inward and, thus, $p+k+q=0$. 
The expression of the one-loop gluon self-energy in FG is well known: 
\be
\Pi^{\mu\nu}_{(FG)}(k)=N g^2 \int dp \frac{1}{p^2 q^2} 
   \biggl[2p^{\mu}p^{\nu}+2q^{\mu}q^{\nu}-
   (p^2+q^2) g^{\mu\nu}-k^{\mu}k^{\nu} 
              +2k^2 d^{\mu\nu} \biggr].
\ee

The one-loop pinch contribution to the gluon self-energy in FG is calculated 
as follows. We consider the $S$-matrix element $T$ for 
the quark-quark scattering at one-loop order.  
Since the gluon propagator in FG does not have a longitudinal  
$k^{\mu}k^{\nu}$ term, the pinch contribution to $T$ only comes from 
the vertex diagram of the second kind with the three-gluon 
vertex of the type $\Gamma^{P}$ (and its mirror graph)~\cite{rCP}, 
and is given by
\be
      T_{P(FG)}^{V_2}=-2 N g^2[T^a \gamma ^{\alpha}]  \int dp
             \frac{1}{p^2 q^2} D_{\alpha \beta}(k) [T^a \gamma ^{\beta}]~.
\ee
The inverse of the gluon propagator is 
\be
      {[}D^{-1}_{(FG)}{]}^{\mu \nu}(k)
        = -k^2 g^{\mu \nu}~,
\ee
and thus $D_{(FG)}$ and its inverse satisfy the identities in 
Eq.(\ref{Identity}). 
We can then apply the formulae Eqs.(\ref{TPV2}) and (\ref{PIV2}) to 
$T_{P(FG)}^{V_2}$ obtaining the FG pinch contribution to the gluon 
self-energy 
\be
  \Pi^{\mu \nu}_{P(FG)}(k) =2N g^2 k^2 d^{\mu \nu}\int dp  
             \frac{1}{p^2 q^2} .
\ee
The sum of $\Pi^{\mu \nu}_{(FG)}$ and $\Pi^{\mu \nu}_{P(FG)}$ is given by 
\be
 \widehat{\Pi}^{\mu \nu} (k) = N g^2 \int dp   
    \frac{1}{p^2q^2}\biggl[2p^{\mu}p^{\nu}+2q^{\mu}q^{\nu}
          -(p^2+q^2) g^{\mu\nu}-k^{\mu}k^{\nu} 
     +4k^2 d^{\mu\nu}\biggr]~.   
\label{InvSelf}
\ee
This is the effective gluon self-energy obtained 
before in the PT framework~\cite{rCa}\cite{rCP}\cite{Papavass}. 
It is noted that $\widehat{\Pi}^{\mu \nu} (k)$ can also be derived 
without using PT but by the background field method with a special 
value of the gauge parameter $\xi_Q=1$~\cite{Hiroshima}.

The effective gluon self-energy $\widehat{\Pi}^{\mu \nu} (k)$ 
has the following features: \\
(i) It satisfies the transversality relation.  
Indeed using $k+p+q=0$ we find 
\bea
    \widehat{\Pi}^{\mu \nu} (k)k_{\nu}&=&2 N g^2 \int dp 
     \Bigl\{\frac{p^{\mu}}{p^2}+\frac{q^{\mu}}{q^2}  \Bigr\} 
     \nonumber \\
   &=&0~.
\label{Transversality}
\eea
(ii) As was shown explicitly at one-loop level~\cite{rCP}, the 
PT modified gluon three-point function 
$g f^{abc}\widehat{\Gamma}_{\mu \nu \alpha}$ and 
$\widehat{\Pi}^{\mu \nu}(k)$    
satisfy the following {\it tree-level} Ward-Takahashi identity 
\be
  p^{\mu}\widehat{\Gamma}_{\mu \nu \alpha}(p,q,r)=
     - \widehat{\Pi}_{\nu \alpha}(q)+\widehat{\Pi}_{\nu \alpha}(r)~. 
\label{WI}
\ee
This implies that the wave function renormalization for 
the PT modified gluon self-energy 
$\widehat{\Pi}_{\mu \nu}$  
contains the running of the QCD couplings. Indeed, at zero temperature,  
after integration and renormalization
it is rewritten  as    
\be
   \widehat{\Pi}^{\mu \nu} (k) = g^2 (g^{\mu \nu} k^2-k^{\mu}k^{\nu})
    \Big(b~ {\rm ln}\frac{k^2}{\mu^2}+ {\rm const} \Big)~,
\label{11N3}
\ee
where $b=-11N/(48\pi^2)$ is the coefficient of $g^3$ in the usual
QCD $\beta$ function without fermions. 

\bigskip

\setcounter{equation}{0}
\section{PT Gluon Self-Energy in the Coulomb Gauge}
\smallskip

The gauge fixing term in the Coulomb gauge (CG) is given by 
\be
   {\cal L}=-\frac{1}{2\xi_C}(\partial^i A^a_i)^2 .
\ee
Then, with a unit vector $n^{\mu}=(1,0,0,0)$, 
the CG gluon propagator,
$iD_{ab(CG)}^{\mu \nu}=i\delta _{ab}D_{(CG)}^{\mu \nu}$,  
and its inverse are expressed as 
\bea
    D_{(CG)}^{\mu \nu}(k)&=&-\frac{1}{k^2}\biggl[g^{\mu \nu}+
     \biggl(1-\xi_C \frac{k^2}{{\bf k}^2} \biggr) 
                \frac{k^{\mu}k^{\nu}}{{\bf k}^2} -
      \frac{k_0}{{\bf k}^2}(k^{\mu}n^{\nu}+n^{\mu}k^{\nu})\biggr] 
   \\
   {[}D^{-1}_{(CG)}{]}^{\mu \nu}(k)&=&-k^2\biggl[g^{\mu \nu}-
        \frac{k^{\mu}k^{\nu}}{k^2}\biggr]+\frac{1}{\xi_C}
        \biggl[k^{\mu}k^{\nu}-k_0(k^{\mu}n^{\nu}+n^{\mu}k^{\nu})+
                         k_0^2 n^{\mu}n^{\nu}\biggr]~.  \nonumber 
\eea 
The three-gluon vertex is the same as in FG, that is, 
$\Gamma^{abc}_{\lambda \mu \nu } (p, k, q)$ in Eq.(\ref{VertexFG}), and the 
ghost propagator $i\delta^{ab}G_{(CG)}$ and the ghost-gluon vertex 
$\Gamma^{abc}_{\mu (CG)}(p,k,q)$ (see Fig.5(a)) in CG 
are given by 
\bea
     G_{(CG)}(k)&=&\frac{1}{{\bf k}^2},   \nonumber   \\
     \Gamma^{abc}_{\mu (CG)}(p,k,q)&=&gf^{abc}[p_{\mu}-p_0 n_{\mu}]  .
\eea

In the limit $\xi_C=0$, 
$D_{(CG)}^{\mu \nu}(k)$ reduces to the well-known form~\cite{Ali}
\be
  D_{(CG)}^{00}=\frac{1}{{\bf k}^2}, \quad 
  D_{(CG)}^{0i}=0, \quad   
  D_{(CG)}^{ij}=\frac{1}{k^2}\biggl(\delta^{ij}-
       \frac{k^i k^j}{{\bf k}^2}\biggr)~.
\label{CGpropagatorlimit} 
\ee
However, its inverse does not exist in this limit. 
The one-loop CG gluon self-energy was calculated in Ref.\cite{HKT} in the  
$\xi_C=0$ limit using the gluon propagator in Eq.(\ref{CGpropagatorlimit}).

In the framework of PT, we need to use the identities in Eq.(\ref{Identity}), 
satisfied by 
the gluon propagator and its inverse, to extract from $T_P$ 
the pinch contributions to 
the gluon self-energy. Therefore, in principle,  we must work 
with a non-zero  $\xi_C$. Thus we recalculate the one-loop 
gluon self-energy in CG with an arbitrary gauge parameter  $\xi_C$. 
The results for the contributions from  
Fig.1(b), the tadpole diagram (Fig.1(c)), and the ghost diagram (Fig.1(d)) 
are respectively as follows:  

\bea
  \Pi^{\mu \nu}_{(a) (CG)}(k) &=& \frac{N}{2} g^2 \int dp 
            \frac{1}{p^2 q^2} \times  \nonumber \\
      & & \times \Biggl[ g^{\mu\nu} \biggl\{ 8k^2 - 
            \biggl[ \biggl(\frac{k^2 (k^2 - 2q^2 -
 4 {\bf k}\cdot {\bf p}) + q^4}{{\bf p}^2} +p^2 \biggr)  
         +   (p \leftrightarrow q)  \biggr] \biggr\} \nonumber \\ 
  & & \quad + \biggl\{ p^{\mu}p^{\nu} \biggl[ -3 + 
     \frac{({\bf p}\cdot {\bf q})^2 }
             {{\bf p}^2 {\bf q}^2} 
     +\frac{4{\bf p}\cdot {\bf q}}{{\bf p}^2}
          \nonumber \\ 
 & &  \qquad \qquad \qquad 
  -\frac{q^2}{{\bf p}^2 {\bf q}^2} 
   (3p^2+2q^2+4{\bf p}^2+{\bf q}^2+4p_0q_0)
 \biggr] + (p \leftrightarrow q)  \biggr\}  \nonumber \\
 & & \quad + ( p^{\mu}q^{\nu}+q^{\mu}p^{\nu} ) 
   \biggl[ -5- \frac{({\bf p}\cdot {\bf q})^2 }
             {{\bf p}^2 {\bf q}^2} 
     +2({\bf p}\cdot {\bf q}) \Bigl( 
           \frac{1}{{\bf p}^2} + \frac{1}{{\bf q}^2} \Bigr)
   -\frac{p^2q^2}{{\bf p}^2 {\bf q}^2}
       \biggr]  \nonumber  \\
  & & \quad + \biggl\{ (n^{\mu}p^{\nu}+p^{\mu}n^{\nu}) 
         \frac{1}{{\bf p}^2 {\bf q}^2} 
     \biggl[ kq \Bigl( p^2 q_0 - q^2 p_0 
            -2 {\bf p}\cdot {\bf q} (p_0-q_0) \Bigr) 
   \nonumber \\
  & & \qquad \qquad \qquad -4k_0 p_0q_0 (pq) 
     -  \Bigl( q^2{\bf p}^2q_0 
       + p^2{\bf q}^2p_0 -{\bf p}\cdot {\bf q}
         (p^2 q_0 + q^2 p_0) \Bigr)  \nonumber \\
     & &  \qquad \qquad  \qquad \qquad  +p^2q^2q_0 +q^2 {\bf q}^2 p_0
 \biggr] +  (p \leftrightarrow q)  \biggr\}  \nonumber \\
 & & \quad + n^{\mu}n^{\nu} \frac{2p_0q_0}
   {{\bf p}^2 {\bf q}^2} (-2k^2(pq) -p^2q^2) \Biggr]   
    \nonumber \\
 &+&\xi_C \frac{N}{2} g^2  \int dp 
      \Biggl[ g^{\mu\nu} \biggl\{ \biggl(-k^4 \frac{1}{q^2{\bf p}^4}
    +k^2\frac{2}{{\bf p}^4}-\frac{q^2}{{\bf p}^4}
            \biggr) + (p \leftrightarrow q)  \biggr\}  \nonumber \\
& & \quad + \biggl\{ p^{\mu}p^{\nu}  \biggl[ 
     \frac{1}{p^2 {\bf p}^2 {\bf q}^4}
     \Bigl( k^2 {\bf p}^2 - (kq)^2 -2(kp)p^2 -2k_0p_0(kq) \Bigr)
        \nonumber  \\
 & &  \qquad  \qquad  + \frac{1}{q^2 {\bf p}^4 {\bf q}^2}
     \Bigl( k^2 {\bf q}^2 - (kq)^2 + 2k_0q_0(kq) \Bigr) 
 - \frac{2}{{\bf p}^4}-\frac{1}{{\bf q}^4}
      \biggr] + (p \leftrightarrow q)  \biggr\}  \nonumber \\
  & & \quad + ( p^{\mu}q^{\nu}+q^{\mu}p^{\nu} )
      \biggl[\biggl(\frac{k^2{\bf q}^2+(kp)(kq)+k_0q_0(p^2-q^2)}
               {q^2{\bf p}^4{\bf q}^2}+
      \frac{kq}{{\bf p}^4{\bf q}^2}-\frac{2}{{\bf p}^4}
    \biggr) + (p \leftrightarrow q)  \biggr] \nonumber \\
  & & \quad + \biggl\{ (n^{\mu}p^{\nu}+p^{\mu}n^{\nu})
         \frac{kq}{{\bf p}^2{\bf q}^2}
   \biggl[k^2  \Bigl( \frac{q_0}{q^2{\bf p}^2} -
                          \frac{p_0}{p^2{\bf q}^2}    \Bigr)
    - \Bigl( \frac{q_0}{{\bf p}^2} -
                          \frac{p_0}{{\bf q}^2}    \Bigr)
    \biggr] + (p \leftrightarrow q)  \biggr\} \Biggr]  \nonumber \\ 
 &+&\xi_C^2 \frac{N}{2} g^2 \int dp 
      \frac{1}{{\bf p}^4 {\bf q}^4} 
           \biggl[ \biggl\{(kq)^2  p^{\mu}p^{\nu} 
              + (p \leftrightarrow q) \biggr\} 
        -(kp)(kq) (p^{\mu}q^{\nu}+q^{\mu}p^{\nu}) \biggr],
\eea

\bea
  \Pi^{\mu \nu}_{(b)(CG)}(k) &=& \frac{N}{2} g^2 \int dp 
    \Biggl[ g^{\mu\nu} \biggl\{\frac{1}{{\bf p}^2}-
          \frac{1}{p^2}+(p \leftrightarrow q) \biggr\} \nonumber \\
& &  \qquad \qquad  
   +\biggl\{ p^{\mu}p^{\nu}\frac{1}{p^2{\bf p}^2}
       +(p \leftrightarrow q) \biggr\}
   +\biggl\{(n^{\mu}p^{\nu}+p^{\mu}n^{\nu})\frac{-p_0}{p^2{\bf p}^2}
        +(p \leftrightarrow q) \biggr\} \Biggr] \nonumber \\
 &+&\xi_C \frac{N}{2} g^2  \int dp 
     \biggl[ g^{\mu\nu} \Bigl(\frac{p^2}{{\bf p}^4}
        +(p \leftrightarrow q)  \Bigr)
      + p^{\mu}p^{\nu} \Bigl(\frac{-1}{{\bf p}^4}
    +(p \leftrightarrow q)  \Bigr) \biggr],        \\
\nonumber \\
  \Pi^{\mu \nu}_{Ghost(CG)}(k) &=& \frac{N}{2} g^2 \int dp 
    \frac{1}{{\bf p}^2{\bf q}^2}  \nonumber \\
 & & \times \biggl[ ( p^{\mu}q^{\nu}+q^{\mu}p^{\nu} ) 
     - \Bigl( q_0(n^{\mu}p^{\nu}+p^{\mu}n^{\nu}) 
             +(p \leftrightarrow q)  \Bigr) + 
        n^{\mu}n^{\nu} 2p_0q_0 \biggr],
\eea
where we have chosen the variables as $ k+p+q=0 $ and, therefore, 
the integrands can be written in the forms which are symmetric 
in the variables $p$ and $q$. Here and in the following, the notation 
$+(p \leftrightarrow q)$ implies symmetrization of the preceding term 
under interchange of $p$ and $q$. 
The one-loop CG gluon self-energy  is given by the sum 
\be
  \Pi^{\mu \nu}_{(CG)} = \Pi^{\mu \nu}_{(a)(CG)}+\Pi^{\mu \nu}_{(b)(CG)}
                              +\Pi^{\mu \nu}_{Ghost(CG)}.
\label{PolaCG}
\ee
We have checked that the $\xi_C$-independent part of
$\Pi^{\mu \nu}_{(CG)}$ agrees 
with the results given in Eqs.(4.6), (4.8), (4.10), and (4.12) of 
Ref.\cite{HKT}.

We now calculate the pinch contributions to the CG gluon self-energy.
Since the CG gluon propagator and its inverse satisfy the relations in 
Eq.(\ref{Identity}), that is, 
\bea
       D^{(CG)}_{\alpha \mu}(k)[D^{-1}_{(CG)}]^{\mu \beta}(k)&=&
    D^{(CG)}_{\alpha \mu}(k)[-k^2d^{\mu \beta}] + 
       \frac{k_{\alpha}}{{\bf k}^2}(k_0 n^{\beta}- k^{\beta})  
\nonumber  \\
       D^{-1}_{(CG)\alpha \mu }(k)D^{\mu \beta}_{(CG)}(k) 
          &=& [-k^2 d_{\alpha \mu}]D^{\mu \beta}_{(CG)}(k)+ 
         (k_0 n_{\alpha}- k_{\alpha})\frac{k^{\beta}}{{\bf k}^2}~, 
\eea
we can follow the prescription explained in Sec.2 to extract 
the one-loop pinch contributions.
The individual contributions in CG from the vertex (first and
second kind) and box diagrams are presented in 
Appendix A.1. In total the pinch contribution to the CG gluon self-energy 
is expressed as
\bea
  \Pi^{\mu \nu}_{P(CG)}(k) &=& \frac{N}{2} g^2  k^2 d^{\mu \nu} \int dp 
     \frac{1}{p^2 q^2} \biggl[\frac{k^2 - q^2 -
 4 {\bf k}\cdot {\bf p}}{{\bf p}^2} 
      +  (p \leftrightarrow q) \biggr]  \nonumber \\
    &+&\frac{N}{2}g^2  k^2 d^{\mu \alpha} d^{\nu \beta}\int dp 
\frac{1}{p^2 q^2 {\bf p}^2 {\bf q}^2}
\biggl\{p_{\alpha}p_{\beta}(k^2 + 4{\bf p}\cdot {\bf q})
           \nonumber \\ 
& &\quad  +(p_{\alpha}n_{\beta}+n_{\alpha}p_{\beta})
      \bigl[p^2q_0-q^2p_0-2{\bf p}\cdot {\bf q}(p_0-q_0)\bigr]
     +n_{\alpha}n_{\beta} 4p_0q_0 (pq) \biggr\} \nonumber \\
&+& \frac{N}{2} g^2 \Biggl[ d^{\mu \alpha} \int dp 
  \biggl\{ p_{\alpha}k^{\nu} \bigl[ \frac{1}{q^2 {\bf p}^2} - 
      \frac{1}{p^2 {\bf q}^2} + 
   \bigl( \frac{1}{q^2}-\frac{1}{p^2} \bigr) 
   \frac{{\bf p}\cdot {\bf q} }
   {{\bf p}^2 {\bf q}^2}  \bigr]  \nonumber \\ 
& &  + n_{\alpha}k^{\nu}\bigl[ 
  - \frac{q_0}{p^2 {\bf q}^2} -\frac{p_0}{q^2 {\bf p}^2}
     +   \bigl(\frac{q_0}{q^2}+\frac{p_0}{p^2} \bigr) 
      \frac{{\bf p}\cdot {\bf q}}
       {  {\bf p}^2 {\bf q}^2}  \bigr] \biggr\}
         + (\mu \leftrightarrow \nu) \Biggr]
          \nonumber \\
 &+&\xi_C \frac{N}{2} g^2  \Biggl[
     k^2  d^{\mu \nu} \int dp 
   \biggl\{k^2 \biggl(\frac{1}{q^2 {\bf p}^4}
         +  \frac{1}{p^2 {\bf q}^4}\biggr) - 
     \frac{1}{{\bf p}^4} -  \frac{1}{{\bf q}^4} 
      \biggr\} \nonumber \\
 & &\quad \quad  + k^2 d^{\mu \alpha} d^{\nu \beta} \int dp 
     \frac{1}{{\bf p}^2 {\bf q}^2} \Biggl\{
   p_{\alpha}p_{\beta}\biggl[k^2 \big( \frac{1}{p^2{\bf q}^2}+
     \frac{1}{q^2{\bf p}^2} \big) - \frac{2}{{\bf p}^2}-
     \frac{2}{{\bf q}^2}\biggr] 
              \nonumber  \\
& &\quad \quad \quad \quad + (p_{\alpha}n_{\beta}+ n_{\alpha}p_{\beta}) 
  \biggl[ \big(\frac{p_0}{{\bf q}^2}-
    \frac{q_0}{{\bf p}^2} \big) - k^2 \big(
   \frac{p_0}{p^2 {\bf q}^2}-\frac{q_0}{q^2 {\bf p}^2}\big)
      \biggr]    \Biggr\}   \nonumber \\
 & &\quad  \quad  + \Biggl\{d^{\mu \alpha} \int dp 
     \frac{p_{\alpha}k^{\nu}}{{\bf p}^2 {\bf q}^2}  
   \bigl(\frac{{\bf k}\cdot {\bf p}}{{\bf q}^2} 
  - \frac{{\bf k}\cdot {\bf q}}{{\bf p}^2} 
     \biggr)   + (\mu \leftrightarrow \nu) 
     \Biggr\}  \Biggr]  \nonumber \\
&+&\xi_C^2 \frac{N}{2} g^2  k^4 d^{\mu \alpha} d^{\nu \beta}\int dp 
\frac{-p_{\alpha}p_{\beta}}{{\bf p}^4 {\bf q}^4}. 
\label{PinchCG}
\eea

In order to compare  the above result with $\Pi^{\mu \nu}_{(CG)}$, 
it is better to express Eq.(\ref{PinchCG}) in terms 
of symmetric tensors  
$g^{\mu\nu}$, $p^{\mu}p^{\nu}$, $q^{\mu}q^{\nu}$, 
$(p^{\mu}q^{\nu}+q^{\mu}p^{\nu})$, $(n^{\mu}p^{\nu}+p^{\mu}n^{\nu})$, 
$(n^{\mu}q^{\nu}+q^{\mu}n^{\nu})$, and $n^{\mu}n^{\nu}$. 
For that purpose, we first write Eq.(\ref{PinchCG}) in terms of  
$g^{\mu\nu}$ and symmetric tensors made up of $k$, $p$ and $n$ and then 
rewrite it in terms of $g^{\mu\nu}$ and symmetric tensors 
made up of $p$, $q$ and $n$. The terms proportional to  
$p^{\mu}p^{\nu}$, $k^{\mu}k^{\nu}$, and 
$(p^{\mu}k^{\nu}+k^{\mu}p^{\nu})$ and to  $(n^{\mu}p^{\nu}+p^{\mu}n^{\nu})$ 
and $(n^{\mu}k^{\nu}+k^{\mu}n^{\nu})$ will then  be rewritten as follows: 
\bea
  & &{\cal R} k^{\mu}k^{\nu}+ {\cal S} (p^{\mu}k^{\nu}+k^{\mu}p^{\nu}) 
  +{\cal T}p^{\mu}p^{\nu} =  \nonumber  \\
  & & \qquad \qquad \qquad  
({\cal R}-2{\cal S}+{\cal T}) p^{\mu}p^{\nu} + 
  ({\cal R}-{\cal S}) (p^{\mu}q^{\nu}+q^{\mu}p^{\nu}) +
   {\cal R}q^{\mu}q^{\nu}  \nonumber \\
  & &{\cal U} (n^{\mu}p^{\nu}+p^{\mu}n^{\nu}) +
     {\cal V} (n^{\mu}k^{\nu}+k^{\mu}n^{\nu})= \nonumber \\
  & & \qquad \qquad \qquad 
    ({\cal U}-{\cal V})(n^{\mu}p^{\nu}+p^{\mu}n^{\nu}) 
        -{\cal V}(n^{\mu}q^{\nu}+q^{\mu}n^{\nu}).
\eea
The final expression for $\Pi^{\mu \nu}_{P(CG)}$ is given in Appendix A.2.

From Eq.(A.4) we find that the one-loop pinch contributions are also 
$\xi_C$-dependent and these $\xi_C$-dependent parts exactly cancel 
against the $\xi_C$-dependent parts of $\Pi^{\mu \nu}_{(CG)}$.
Furthermore it is easy to see that adding the $\xi_C$-independent parts 
of $\Pi^{\mu \nu}_{(CG)}$ and $\Pi^{\mu \nu}_{P(CG)}$,  we obtain 
\bea
  \widetilde \Pi^{\mu \nu}(k)&=& \Pi^{\mu \nu}_{(CG)}(k)+
             \Pi^{\mu \nu}_{P(CG)}(k)\nonumber \\ 
     &=& N g^2 \int dp \frac{1}{p^2q^2}\biggl[ 
           (4k^2-p^2-q^2)g^{\mu \nu} 
                     -3(p^{\mu}p^{\nu}+q^{\mu}q^{\nu})
       -5(p^{\mu}q^{\nu}+q^{\mu}p^{\nu})\biggr]  \nonumber  \\
\label{PITILDE}
\eea
which is equivalent to $\widehat{\Pi}^{\mu \nu} (k)$ in Eq.(\ref{InvSelf}).
Thus we have shown explicitly that the CG gluon self-energy  
$\Pi^{\mu \nu}_{(CG)}$ and the pinch contribution  $\Pi^{\mu \nu}_{P(CG)}$, 
when combined, give the {\it universal} effective gluon self-energy  
$\widehat{\Pi}^{\mu \nu} (k)$. 

We now examine the structure of $\Pi^{\mu \nu}_{P(CG)}$ and discuss some of 
the properties of the CG gluon self-energy itself. 
Only the $\xi_C$-independent parts will be considered.
First, $\Pi^{\mu \nu}_{P(CG)}$ is not transverse. In fact, we easily obtain 
from Eq.(\ref{PinchCG}) 
\bea
\Pi^{\mu \nu}_{P(CG)}k_{\nu}&=& \frac{N}{2} g^2 k^2
    d^{\mu \alpha} \int dp 
    \biggl\{ p_{\alpha}\biggl[ \frac{1}{q^2  {\bf p}^2} - 
      \frac{1}{p^2  {\bf q}^2} + 
   \bigl( \frac{1}{q^2}-\frac{1}{p^2} \bigr) 
   \frac{{\bf p}\cdot {\bf q} }
   { {\bf p}^2  {\bf q}^2}  \biggr]  \nonumber \\ 
& & \qquad  + n_{\alpha}\biggl[ 
  - \frac{q_0}{p^2 {\bf q}^2} -\frac{p_0}{q^2 {\bf p}^2}
     +   \bigl(\frac{q_0}{q^2}+\frac{p_0}{p^2} \bigr) 
      \frac{{\bf p}\cdot {\bf q}}
       {  {\bf p}^2 {\bf q}^2}  \biggr] \biggr\}~,
\eea  
where we have used  $d^{\mu\nu}k_{\nu}=0$ and $d^{\nu\beta}k_{\nu}=0$.
Since the sum 
$\widehat{\Pi}^{\mu \nu} (k)$ satisfies the transversality relation 
(see Eq.(\ref{Transversality})), 
this means that the CG gluon self-energy is not 
transverse either, i.e., 
$\Pi^{\mu \nu}_{(CG)}k_{\nu} \ne 0$, which was indeed pointed out 
in Ref.\cite{HKT}.

Next, let us analyze $\Pi^{\mu \nu}_{P(CG)}$ in the context of hot QCD. 
The hard thermal loop $\delta \Pi^{\mu \nu}$ in the gluon self-energy 
$\Pi^{\mu \nu}$ is the piece proportional to $T^2$, which is 
the leading term in the high-temperature expansion ($T>>\vert {\bf k}\vert$ 
and $T>>\vert k_0 \vert$) and is generated by a small part of the 
integration region in one-loop diagrams with hard momenta of 
order $T$~\cite{BP}. 
It is known that $\delta \Pi^{\mu \nu}$ is 
gauge independent and satisfies the transversality relation 
(the Ward Identity) $k^{\mu}\delta \Pi^{\mu \nu}(k)=0$~\cite{HardLoop}. 
Now $\Pi^{\mu \nu}$ has a dimension of ${\rm mass}^2$ and, apart from the 
tensorial factors, is composed of  non-singular functions of 
$\vert {\bf k}\vert$ and $k_0$. 
Then look at the structure of  $\Pi^{\mu \nu}_{P(CG)}$ in 
Eq.(\ref{PinchCG}) (see also $\xi_C$-dependent parts). 
It is made up of the terms proportional to $k^2d^{\mu\nu}$, 
$k^2d^{\mu\alpha}d^{\nu\beta}$ and $d^{\mu\alpha}k^{\nu}$. 
The $d^{\mu\alpha}k^{\nu}$ terms appear as a result of using the 
formula in Eq.(\ref{Formula}), that is, 
$k^2d^{\mu\alpha}p_j^{\nu}=k^2d^{\mu\alpha}d^{\nu\tau}p_{j\tau}+
d^{\mu\alpha}k^{\nu}(kp_j)$. 
So by a simple dimensional analysis, we easily see that there is no way 
for the pinch contribution $\Pi^{\mu \nu}_{P(CG)}$ to produce a $T^2$-term. 
This means that $\Pi^{\mu \nu}_{P(CG)}$ does not contribute to the 
hard thermal loop $\delta \Pi^{\mu \nu}$. These arguments
can be applied to the pinch contributions to the gluon self-energy 
calculated in any gauge (see Sec.5 for the TAG calculation). 
It is clear from the discussion in Sec.2 that by construction, 
the terms in the pinch parts always carry such factors as 
$k^2d^{\mu\nu}$, $k^2d^{\mu\alpha}$, $k^4d^{\mu\nu}$ and
$k^4d^{\mu\alpha}d^{\nu\beta}$, and hence they do not generate a $T^2$-term. 
The gluon self-energy calculated in any gauge, when combined with  the 
pinch contribution, gives the universal and thus {\it gauge-independent} 
$\widehat{\Pi}^{\mu \nu} (k)$. As the pinch part does not contribute to 
the hard thermal loop $\delta \Pi^{\mu \nu}$, 
$\delta \Pi^{\mu \nu}$ should be  gauge-independent. 
Moreover $\delta \Pi^{\mu \nu}$ should satisfy the transversality relation 
$k^{\mu}\delta \Pi^{\mu \nu}(k)=0$ since $\widehat{\Pi}^{\mu \nu} (k)$ does.
This is an explanation for the gauge-independence 
and the transverse nature of the hard thermal loop 
$\delta \Pi^{\mu \nu}$ from the PT point of view.

In a similar way we can argue for the gauge-independence of the  
electric mass $m_{el}$  and 
``effective gluon mass'' $m_G$ in hot QCD.
From the expression of $\Pi^{\mu \nu}_{P(CG)}$ in Eq(\ref{PinchCG}), 
we see that its $(00)$-component at $k_0=0$, 
$\Pi^{00}_{P(CG)}(k_0=0, \vert {\bf k}\vert)$,  
vanishes in the limit $\vert {\bf k}\vert \rightarrow 0$. This is true 
for the one-loop pinch contributions calculated in any gauge, 
since, by construction, $\Pi^{00}_{P}$'s (more generally 
$\Pi^{\mu \nu}_{P}$) are proportional to $k^2$. 
Thus
\be
      \lim_{\vert {\bf k}\vert \to 0}  
          \Pi^{00}_{P}(k_0=0, \vert {\bf k}\vert) =0~. 
\ee
On the other hand the limit $\vert {\bf k}\vert \rightarrow 0$ of 
$\Pi^{00}_{(CG)}(k_0=0,\vert {\bf k}\vert)$ remains finite. Hence the limit  
\bea
     \lim_{\vert {\bf k}\vert \to 0}  
           \Pi^{00}_{(CG)}(k_0=0, \vert {\bf k}\vert) 
              &=&\frac{1}{3}Ng^2T^2 \nonumber \\
              &=& m^2_{el}
\eea
is a gauge-independent quantity.  The inverse of 
electric mass $m_{el}$ represents the screening length for static 
electric fields. Another example is provided by
a combination of pinch contributions 
$(1/2)\bigl((k^2/{\bf k}^2)\Pi_{P}^{00}-
\Pi_{P}^{\mu\nu}g_{\mu\nu} \bigr)$ calculated in any gauge.
Obviously the combination is proportional to 
$k^2$ and thus its limit as  $k^2 \rightarrow 0$ is 0. 
Therefore, the limit 
\bea
     m_G^2&=& \lim_{k^2\to0} \frac{1}{2}\biggl( 
              \frac{k^2}{{\bf k}^2}\Pi_{(CG)}^{00} -
                \Pi_{(CG)}^{\mu\nu}g_{\mu\nu} \biggr)   \nonumber \\
         &=&\frac{1}{6}Ng^2T^2~,  
\eea
is a gauge independent quantity and is called  ``effective gluon mass'' 
squared.

\bigskip

\setcounter{equation}{0}
\section{PT Gluon Self-Energy in the Temporal Axial Gauge}
\smallskip

The gauge fixing term in the temporal axial gauge (TAG) is provided by 
\be
   {\cal L}=-\frac{1}{2\xi_A}(n^{\mu} A^a_{\mu})^2 ,
\ee
where $n^{\mu}=(1,0,0,0)$. The gluon propagator in TAG,    
$iD_{ab(TAG)}^{\mu \nu}=i\delta _{ab}D_{(TAG)}^{\mu \nu}$, and its 
inverse are given by
\bea
    D_{(TAG)}^{\mu \nu}(k)&=&-\frac{1}{k^2}\biggl[g^{\mu \nu}+
     (1+\xi_A k^2) 
            \frac{k^{\mu}k^{\nu}}{k_0^2} - 
 \frac{1}{k_0}(k^{\mu}n^{\nu}+n^{\mu}k^{\nu})\biggr] 
           \label{TagProp}    \\
   {[}D^{-1}_{(TAG)}{]}^{\mu \nu}(k)&=&
      -k^2\biggl(g^{\mu \nu}-\frac{k^{\mu}k^{\nu}}{k^2}\biggr) -
     \frac{1}{\xi_A}n^{\mu}n^{\nu}.
\eea 
It is noted that the gauge parameter $\xi_A$ in TAG has a dimension of 
${\rm mass}^{-2}$.
The three-gluon vertex is given again by 
$\Gamma^{abc}_{\lambda \mu \nu } (p, k, q)$ in Eq.(\ref{VertexFG}), 
and the ghost propagator $i\delta^{ab}G_{(TAG)}$ and the ghost-gluon vertex 
$\Gamma^{abc}_{\mu (TAG)}(p,k,q)$ (see Fig.5(b)) in TAG
are, respectively, 
\bea
     G_{(TAG)}(k)&=&\frac{-i}{k_0},   \nonumber   \\
     \Gamma^{abc}_{\mu (TAG)}(p,k,q)&=&gf^{abc}[-i n_{\mu}]  .
\eea
 
The one-loop gluon self-energy in TAG was calculated in 
Refs.\cite{Hot}\cite{HKT} in the $\xi_A=0$ limit. There, the ghost loop 
contribution was omitted due to the argument 
that the ghost field decouples in this limit.
However, in the limit $\xi_A=0$ the inverse of the gluon 
propagator does not exist.
So in the framework of PT we work with a non-zero $\xi_A$. 
We recalculate the gluon self-energy using the gluon propagator with an 
arbitrary $\xi_A$ given in Eq.(\ref{TagProp}). 
For a non-zero $\xi_A$ the ghost should 
be taken into account and at one-loop level it contributes to 
the $\xi_A$-independent part of $\Pi^{00}_{(TAG)}$~\cite{HKT}.
The contributions of 
Fig.1(b), of the tadpole diagram (Fig.1(c)), and the ghost diagram (Fig.1(d)) 
are, respectively, 
\bea
  \Pi^{\mu \nu}_{(a) (TAG)}(k) &=& \frac{N}{2} g^2 \int dp 
                  \frac{1}{p^2 q^2}   \nonumber \\
  & & \times \Biggl[ g^{\mu\nu} \biggl\{ 8k^2 - 
            \biggl[ \biggl(\frac{k^2 (k^2 +2p^2 - q^2 -
 4 {\bf k}\cdot {\bf p}) -p^2q^2}{p_0^2} +p^2 \biggr)  
         +   (p \leftrightarrow q)  \biggr] \biggr\} \nonumber \\ 
  & & \quad + \biggl\{ p^{\mu}p^{\nu} \biggl[ -3 + 
     \frac{({\bf p}\cdot {\bf q})^2 }{p_0^2 q_0^2}
   - \frac{ 2{\bf p}\cdot {\bf q}}{p_0^2 q_0^2}
         (q^2-2q_0^2) +\frac{q^4}{p_0^2 q_0^2}    \nonumber \\ 
 & &  \qquad \qquad \qquad \qquad \qquad \qquad \qquad \qquad
  -\frac{p^2+q^2}{p_0^2}+ \frac{q^2}{q_0^2}
 \biggr] + (p \leftrightarrow q)  \biggr\}  \nonumber \\
 & & \quad + ( p^{\mu}q^{\nu}+q^{\mu}p^{\nu} ) 
   \biggl[ -5- \frac{({\bf p}\cdot {\bf q})^2 }
             {p_0^2 q_0^2} 
     +({\bf p}\cdot {\bf q}) \Bigl(
        \frac{p^2+q^2}{p_0^2 q_0^2} 
          + \frac{2}{p_0^2} + \frac{2}{q_0^2} \Bigr)
   -\frac{p^2q^2}{p_0^2 q_0^2} \biggr]  \nonumber  \\
 & & \quad + \biggl\{ (n^{\mu}p^{\nu}+p^{\mu}n^{\nu}) 
         \frac{1}{p_0^2 q_0^2} 
     \biggl[ kq \Bigl( -p^2 q_0 + q^2 p_0 
            -2 {\bf p}\cdot {\bf q} (p_0-q_0) \Bigr) 
   \nonumber \\
  & & \qquad \qquad \qquad \qquad \qquad \qquad \qquad -4k_0 p_0q_0 (pq) 
    + q^2p_0q_0^2  \biggr] +  (p \leftrightarrow q)  \biggr\}  \nonumber \\
  & & \quad + n^{\mu}n^{\nu} \frac{2p_0q_0}
   {p_0^2 q_0^2} (-2k^2(pq) -p^2q^2) \Biggr]   
    \nonumber \\
 &+&\xi_A \frac{N}{2} g^2  \int dp 
      \Biggl[ g^{\mu\nu} \biggl\{ \biggl(k^4 \frac{1}{q^2 p_0^2}
    -k^2\frac{2}{p_0^2}+\frac{q^2}{p_0^2}
            \biggr) + (p \leftrightarrow q)  \biggr\}  \nonumber \\
& & \quad + \biggl\{ p^{\mu}p^{\nu}  \biggl[ 
    -k^2 \Bigl(\frac{1}{p^2q_0^2}+ \frac{1}{q^2p_0^2}  \Bigr)
    +\frac{(kq)^2}{p_0^2 q_0^2}\Bigl(\frac{1}{p^2}+ \frac{1}{q^2} \Bigr)
         \nonumber  \\
 & &  \qquad  \qquad \qquad \qquad \qquad - \frac{2k_0(kq)}{p_0^2 q_0^2}
     \Bigl( \frac{q_0}{q^2}-\frac{p_0}{p^2} \Bigr) 
    +\frac{2}{p_0^2}+\frac{1}{q_0^2}
      \biggr] + (p \leftrightarrow q)  \biggr\}  \nonumber \\
  & & \quad + ( p^{\mu}q^{\nu}+q^{\mu}p^{\nu} )
    \biggl[\biggl(\frac{-k^2p_0^2-(kp)(kq)+k_0p_0(p^2-q^2)}{p^2p_0^2q_0^2}+
      \frac{2}{p_0^2}
    \biggr) + (p \leftrightarrow q)  \biggr] \nonumber \\
 & & \quad + \biggl\{ (n^{\mu}p^{\nu}+p^{\mu}n^{\nu})
         \frac{kq}{p_0^2q_0^2}
   \biggl[k^2  \Bigl( \frac{p_0}{p^2} - \frac{q_0}{q^2}    \Bigr)
 +q_0-p_0 \biggr] + (p \leftrightarrow q)  \biggr\} \Biggr]  \nonumber \\ 
 &+&\xi_A^2 \frac{N}{2} g^2 \int dp 
      \frac{1}{p_0^2 q_0^2} 
           \biggl[ \biggl\{(kq)^2  p^{\mu}p^{\nu} 
              + (p \leftrightarrow q) \biggr\} 
        -(kp)(kq) (p^{\mu}q^{\nu}+q^{\mu}p^{\nu}) \biggr]
\eea

\bea
  \Pi^{\mu \nu}_{(b) (TAG)}(k) &=& \frac{N}{2} g^2 \int dp 
    \Biggl[ g^{\mu\nu} \biggl\{\frac{-1}{p^2}+
          \frac{-1}{p_0^2}+(p \leftrightarrow q) \biggr\} \nonumber \\
& &  \qquad \qquad  
   +\biggl\{ p^{\mu}p^{\nu}\frac{1}{p^2p_0^2}
       +(p \leftrightarrow q) \biggr\}
   +\biggl\{(n^{\mu}p^{\nu}+p^{\mu}n^{\nu})\frac{-1}{p^2p_0}
        +(p \leftrightarrow q) \biggr\} \Biggr] \nonumber \\
 &+&\xi_A \frac{N}{2} g^2  \int dp 
     \biggl[ g^{\mu\nu} \Bigl(\frac{-p^2}{p_0^2}
        +(p \leftrightarrow q)  \Bigr)
      + p^{\mu}p^{\nu} \Bigl(\frac{1}{p_0^2}
    +(p \leftrightarrow q)  \Bigr) \biggr]  \nonumber  \\
 \nonumber \\
  \Pi^{\mu \nu}_{Ghost (TAG)}(k) &=& \frac{N}{2} g^2 \int dp 
         n^{\mu}n^{\nu}\frac{2}{p_0q_0}. 
\eea
The one-loop gluon self-energy  in TAG is then given by the sum 
\be
  \Pi^{\mu \nu}_{(TAG)} = \Pi^{\mu \nu}_{(a)(TAG)}+\Pi^{\mu \nu}_{(b)(TAG)}
                              +\Pi^{\mu \nu}_{Ghost(TAG)}.
\label{PolaTAG}
\ee
The $\xi_A$-independent part of $\Pi^{\mu \nu}_{(TAG)}$ agrees with 
the results given in Eqs.(4.5), (4.7), (4.9), (4.11) of Ref.\cite{HKT} 
except for the ghost contribution to $\Pi^{00}_{(TAG)}$. 

We now calculate the pinch contributions in TAG.
Since the TAG propagator and its inverse 
satisfy the relations in Eq.(\ref{Identity}), i.e.,
\bea
       D^{(TAG)}_{\alpha \mu}(k)[D^{-1}_{(TAG)}]^{\mu \beta}(k)&=&
    D^{(TAG)}_{\alpha \mu}(k)[-k^2d^{\mu \beta}] + 
       k_{\alpha}\biggl(\frac{n^{\beta}}{k_0}-\frac{k^{\beta}}{k^2}\biggr) 
  \nonumber  \\
       D^{-1}_{(TAG)\alpha \mu }(k)D^{\mu \beta}_{(TAG)}(k) 
          &=& [-k^2 d_{\alpha \mu}]D^{\mu \beta}_{(TAG)}(k)+ 
         \biggl(\frac{n_{\alpha}}{k_0}-\frac{k_{\alpha}}{k^2}\biggr)
         k^{\beta},
\eea
we can follow the same procedure as before and we obtain for the 
pinch contribution to the gluon self-energy in TAG,

\bea
  \Pi^{\mu \nu}_{P(TAG)}(k) &=& \frac{N}{2} g^2  k^2 d^{\mu \nu} \int dp 
     \frac{1}{p^2 q^2} \biggl[\frac{k^2 +2p^2-  q^2 -
 4 {\bf k}\cdot {\bf p}}{p_0^2} 
      +  (p \leftrightarrow q) \biggr]  \nonumber \\
    &+&\frac{N}{2}g^2  k^2 d^{\mu \alpha} d^{\nu \beta}\int dp 
\frac{1}{p^2 q^2 p_0^2 q_0^2}
\biggl\{p_{\alpha}p_{\beta}(4p_0q_0-k^2)
           \nonumber \\ 
& &\quad  +(p_{\alpha}n_{\beta}+n_{\alpha}p_{\beta})
      \bigl[-p^2q_0+q^2p_0-2{\bf p}\cdot {\bf q}(p_0-q_0)\bigr]
     +n_{\alpha}n_{\beta} 4p_0q_0 (pq) \biggr\} \nonumber \\
 &+&\xi_A \frac{N}{2} g^2  \Biggl[
     k^2  d^{\mu \nu} \int dp 
   \biggl\{-k^2 \biggl(\frac{1}{q^2 p_0^2}
         +  \frac{1}{p^2 q_0^2}\biggr) +
     \frac{1}{p_0^2} + \frac{1}{q_0^2} 
      \biggr\} \nonumber \\
 & &\quad \quad  + k^2 d^{\mu \alpha} d^{\nu \beta} \int dp 
     \frac{1}{p_0^2 q_0^2} \Biggl\{
   p_{\alpha}p_{\beta}\biggl[-k^2 \big( \frac{1}{p^2}+
     \frac{1}{q^2} \big) \biggr] 
              \nonumber  \\
& &\quad \quad \quad \quad \qquad + (p_{\alpha}n_{\beta}+ n_{\alpha}p_{\beta}) 
  \biggl[ q_0-p_0 +k^2 \big(
   \frac{p_0}{p^2}-\frac{q_0}{q^2}\big)
      \biggr]    \Biggr\}     \Biggr]  \nonumber \\
&+&\xi_A^2 \frac{N}{2} g^2  k^4 d^{\mu \alpha} d^{\nu \beta}\int dp 
\frac{-p_{\alpha}p_{\beta}}{p_0^2q_0^2}. 
\label{PinchTAG}
\eea
The individual contributions in TAG from the vertex (first and second kind) 
and box diagrams are presented in Appendix (B.1).

The expression of $\Pi^{\mu \nu}_{P(TAG)}$ is further rewritten in terms of 
symmetric tensors  
$g^{\mu\nu}$, $p^{\mu}p^{\nu}$, $q^{\mu}q^{\nu}$, 
$(p^{\mu}q^{\nu}+q^{\mu}p^{\nu})$, $(n^{\mu}p^{\nu}+p^{\mu}n^{\nu})$, 
$(n^{\mu}q^{\nu}+q^{\mu}n^{\nu})$, and $n^{\mu}n^{\nu}$. The result is given  
in Appendix (B.2).  From this expression 
we can see that the one-loop pinch contributions are also 
$\xi_A$-dependent and these $\xi_A$-dependent parts exactly cancel 
against the $\xi_A$-dependent parts of $\Pi^{\mu \nu}_{(TAG)}$.
Also we find that the sum of $\Pi^{\mu \nu}_{(TAG)}$ and 
$\Pi^{\mu \nu}_{P(TAG)}$ is equal to $\widetilde \Pi^{\mu \nu}$ in 
Eq.(\ref{PITILDE}) and thus equal to the {\it universal} 
$\widehat \Pi^{\mu \nu}$ in Eq.(\ref{InvSelf}). 

Let us now examine the results of these TAG calculations. 
We will only consider the $\xi_A$-independent part. 
First it is easily seen from Eq.(\ref{PinchTAG}) that the pinch contribution 
$\Pi^{\mu \nu}_{P(TAG)}$ is transverse, i.e., 
$k_{\mu}\Pi^{\mu \nu}_{P(TAG)}=0$. 
Hence the TAG gluon self-energy $\Pi^{\mu \nu}_{(TAG)}$ should be 
transverse~\cite{HKT}. 
Here it is noted that we have included the ghost-loop contribution
in $\Pi^{\mu \nu}_{(TAG)}$.

At zero temperature ($T=0$) the $\xi_A$-independent part of the
pinch contribution 
$\Pi^{\mu \nu}_{P(TAG)}$ does not contain ultraviolet divergences. 
This can be easily seen from the examination of the $g^{\mu\nu}$ part of
$\Pi^{\mu \nu}_{P(TAG)}$ in the limit $\bf k =0$
(and remains true for $\bf k \neq 0$). 
Applying the projection operator $\frac{1}{3}d_{\mu\nu}$ 
to the $\xi_A$-independent part of $\Pi^{\mu \nu}_{P(TAG)}$, 
we find, in the limit $\bf k =0$,
\be
 {\rm d.\ p.\ of}\ \biggl[ \frac{1}{3}d_{\mu\nu}\Pi^{\mu \nu}_{P(TAG)} 
        \biggr]_{\xi_A=0} \!\!\!\!\!\!\!\!
  = {\rm d.\ p.\ of}\ \Biggr[ \frac{N}{6}g^2 k^2_0 \int dp  
    \biggl(\frac{4}{p^2q^2}+\frac{2}{q^2 p^2_0}\biggr) \Biggr]
\ee
where an abbreviation ``${\rm d.\ p.\ of}$" stands for ``divergent part of" and  
we have dropped the $k^2_0$ terms in the numerator of the integrand 
which would only contribute to the finite part. Also we have 
replaced $q_0$ with $-p_0$ and $q$ in the numerator with $-p$, 
since $q=-p-k$ and these replacements do not modify the 
ultraviolet divergent part. As a final step we use the following 
two integral formulae~\cite{CL}:
\bea
  {\rm d.\ p.\ of}\ \ \biggl[ \int dp \frac{1}{p^2(p+k)^2}  
          \biggr] &=& \Delta  \\
   {\rm d.\ p.\ of}\ \ \biggl[ \int dp \frac{1}{(p+k)^2p_0^2}  
        \biggr] &=& -2\Delta,
\eea
where the loop integral $\int dp$ is defined in Eq.(\ref{loop}) and 
$\Delta=\frac{1}{16\pi^2}\frac{2}{4-D}$. Thus we find 
\be
  {\rm d.\ p.\ of}\ \biggl[ 
        \frac{1}{3}d_{\mu\nu}\Pi^{\mu \nu}_{P(TAG)} 
        \biggr]_{\xi_A=0} \!\!\!\!\!\!\!\! =0~,
\ee 
and hence the $\xi_A$-independent part of $\Pi^{\mu \nu}_{P(TAG)}$ is 
ultraviolet finite. 
We have shown in Sec.3 that, at zero temperature, the divergent part of the  
{\it universal} gluon self-energy $\widehat \Pi^{\mu \nu}$, 
which is the sum of $\Pi^{\mu \nu}_{(TAG)}$ and $\Pi^{\mu \nu}_{P(TAG)}$, 
gives us complete information on the correct running of the QCD 
coupling constant at one-loop level. The fact that 
$\Pi^{\mu \nu}_{P(TAG)}$ is ultraviolet finite, therefore, implies that 
in one-loop TAG calculations the only
knowledge of the gluon self-energy is enough to determine the QCD $\beta$
function, which is indeed true for $\xi_A=0$~\cite{CL}.

There is one subtlety in the quantization of gauge theories 
in TAG~\cite{Noncov}\cite{LeibbrandtA}. Spurious  singularities 
appear in the loop-calculations. The gauge condition $n^{\mu} A^a_{\mu}=0$ 
in TAG is not enough to fix the gauge uniquely and there 
still remains a freedom of 
time-independent gauge transformations. This residual invariance manifests 
itself as  unphysical poles 
in the longitudinal part of the gluon propagator given in Eq.(\ref{TagProp}).
In the TAG calculation of the gluon self-energy, these unphysical 
poles in the gluon propagator give  spurious singularities. To 
circumvent these singularities, several methods have been 
proposed, and most noticeable are the 
the principal-value 
prescription~\cite{PV}, the $n^{*}_{\mu}$-prescription~\cite{LeibbrandtB} 
and  the $\alpha$-prescription~\cite{Landshoff}. 

We now know that the longitudinal part of the TAG propagator 
gives rise to pinch parts. Thus the spurious singularities 
due to the unphysical poles of the propagator also appear 
in the pinch contribution. Once this pinch contribution is added to 
the TAG gluon self-energy, the singularities due to 
the ill-fated unphysical poles cancel out. 
To illustrate how these cancellations actually occur, we 
present the PT calculation in TAG of the gauge-independent
thermal $\beta$ function in a hot Yang-Mills gas~\cite{Sasakia}.  
 
As stated in Sec.3, the PT modified gluon self-energy 
$\widehat\Pi_{\mu \nu}$ contains the running of the QCD coupling.
When the renormalization condition of 
the three-gluon vertex is chosen at the static and symmetric point, 
the thermal $\beta$ function $\beta_T$ is obtained through a formula 
~\cite{BFM}-\cite{EK} 
\be
   \beta_T \equiv T\frac{dg(T,\kappa)}{dT}=\frac{g}{2\kappa^2}
              T \frac{d\Pi_{\perp} (T,\kappa)}{dT},
\label{Beta}
\ee
where $\Pi_{\perp} (T,\kappa)=\Pi_{\perp}
(T,k_0=0,\kappa=\vert  {\bf k} \vert)$ is 
the transverse function of the gluon self-energy 
$\Pi_{\mu \nu}$ at the static limit. Here for $\Pi_{\mu \nu}$ we should use  
$\widehat\Pi_{\mu \nu}$, namely, the sum of the usual one-loop gluon 
self-energy and the pinch contribution. 

In the static limit $k_0=0$, we have 
$\Pi_{\perp} (T,\kappa)=\frac{1}{2}\Pi_{ii}(k_0=0, \kappa)$. The TAG 
calculation of $\Pi_{ii}(k_0=0, \kappa)$ was performed in Ref.\cite{HKT}. 
After the $p_0$ summation and the angular integration, but before the 
$p(=\vert  {\bf p} \vert)$-integration, 
$\Pi^{(TAG)}_{ii}(0, \kappa)$ is given in Eq.(4.43) of Ref.\cite{HKT} as 
\bea
 \Pi_{ii}^{(TAG)}(0,\kappa)&=& \frac{Ng^2}{2\pi^2}\int_{0}^{\infty} dp~p~n(p)
     \nonumber \\ 
   & & \times \Biggl[-2+\frac{\kappa^2}{p^2_{\pm}}+
   \frac{\kappa^4}{4p^2p^2_{\pm}}+\biggl(\frac{2p}{\kappa}+\frac{5\kappa}{2p}-
   \frac{\kappa^3}{2pp^2_{\pm}}-\frac{\kappa^5}{16p^3p^2_{\pm}}\biggr)
       ~{\rm ln} 
        \Big\vert \frac{2p + \kappa}{2p - \kappa} \Big\vert 
      \Biggr]~, \nonumber  \\
\eea
where $n(p)=1/[{\rm exp} (p/T) -1 ]$ is the Bose-Einstein statistical 
distribution function, and 
the principal value prescription was supposed to be applied for 
$1/p^2_{\pm}$. If we do not use the principal value 
prescription and replace $p^2_{\pm}$ with $p^2$, we see that  
the integrand (the terms in $[\cdots]$) would behave as $-4\kappa^2/3p^2$ for 
small $p$. 

Now let us calculate the pinch contribution to $\Pi_{\perp}(T,\kappa)$ 
in TAG.  Applying the projection operator
\be
           t_{ij}=\frac{1}{2}(\delta_{ij}-\frac{k_i k_j}{{ {\bf k}}^2}) 
\ee
to the spatial part of $\Pi^{\mu \nu}_{P(TAG)}$ in 
Eq.(\ref{PinchTAG}) (we are only interested in the $\xi_A$-independent part), 
we obtain in the static limit, 
\bea
     \Pi_{\perp}^{P(TAG)}(T,\kappa)&=&t_{ij}\Pi^{ij}_{P(TAG)}(k_0=0, \kappa) 
                   \nonumber \\
      &=& - N g^2 {\kappa}^2 
   \int dp \biggl\{ 
   \frac{{ {\bf k}}^2 + 4{ {\bf k}} \cdot { {\bf p}}}
        {p^2 q^2 p_0^2} + \frac{1}{p^2  p_0^2} -  \frac{2}{q^2 p_0^2} \biggr\} 
       \nonumber   \\
     & & - \frac{N}{4}g^2 {\kappa}^2 \int dp 
      \biggl[  {\bf p}^2-\frac{( {\bf k}\cdot  {\bf p})^2}
           { {\bf k}^2} \biggr]
    \biggl\{ \frac{ {\bf k}^2}{p^2 q^2 p_0^2 q_0^2} - 
        \frac{4}{p^2 q^2 p_0^2}  \biggr\},
\label{PiPTAG}
\eea
where the terms proportional to $(p_{\alpha}n_{\beta}+n_{\alpha}p_{\beta})$ 
and $n_{\alpha}n_{\beta}$ in $\Pi^{\mu \nu}_{P(TAG)}$ do not contribute 
to $\Pi_{\perp}^{P(TAG)}$. 
After the $p_0$-summation and the angular integration, 
$\Pi_{\perp}^{P(TAG)}(T,\kappa)$ is rewritten as 
\bea
 \Pi_{\perp}^{P(TAG)}(T,\kappa)&=& 
           \frac{Ng^2}{4\pi^2}\int_{0}^{\infty} dp~p~n(p)
     \nonumber \\ 
   & & \times \Biggl[-\frac{\kappa^2}{p^2}-
   \frac{\kappa^4}{4p^4}+\biggl(\frac{\kappa}{p}
   +\frac{\kappa^3}{2p^3}+\frac{\kappa^5}{16p^5}\biggr)
       ~{\rm ln} 
        \Big\vert \frac{2p + \kappa}{2p - \kappa} \Big\vert 
      \Biggr]~,  \nonumber  \\
\eea
where we have used formulae given in Appendix C.
Note that the integrand behaves as $4\kappa^2/3p^2$ for small $p$. 
When $\Pi_{\perp}^{(TAG)}$ and $ \Pi_{\perp}^{P(TAG)}$ are combined 
(remember $\Pi_{\perp}^{(TAG)}=\frac{1}{2}\Pi_{ii}^{(TAG)}(0,\kappa)$), 
the $\kappa^2/p^2$ singularities cancel and the integrand becomes regular as 
$p \rightarrow 0$.  We can, therefore, evaluate the sum 
\bea
  \Pi_{\perp}(T,\kappa)&=&\Pi_{\perp}^{(TAG)}(T,\kappa)
          +\Pi_{\perp}^{P(TAG)}(T,\kappa) \nonumber \\
   &=& \frac{Ng^2}{4\pi^2}\int_{0}^{\infty} dp~p~n(p)
        \Biggl[-2+\biggl(\frac{2p}{\kappa}+ \frac{7\kappa}{2p} \biggr)
       ~{\rm ln}  \Big\vert \frac{2p + \kappa}{2p - \kappa} \Big\vert 
      \Biggr]~,        
\label{InvPi}
\eea
without recourse to the principal value prescription or 
to the other prescriptions mentioned before and 
obtain in the limit  $\kappa << T$ 
\be
   \Pi_{\perp}(T,\kappa) \approx Ng^{2}{\kappa}T \frac{7}{16} 
      + {\cal O} (\kappa^2)~.
\label{InvPiApprox}
\ee
Inserting the above expression into Eq.(\ref{Beta}), we find 
for the gauge-independent thermal $\beta$ function 
\be
      \beta_T=g^3 N\frac{7}{32}\frac{T}{\kappa}, 
\label{InvBeta} 
\ee
which coincides with the result of Refs.\cite{ACPS}\cite{Sasakia}. 
What we have learned from these calculations is that 
spurious singularities in TAG appear only in the gauge-dependent 
parts and that when we deal with physical and/or gauge-independent quantities, 
these singularities cancel among themselves and disappear.  

\bigskip

\setcounter{equation}{0}
\section{Summary and Discussion}
\smallskip

In this paper we have used the $S$-matrix PT and calculated the one-loop 
effective gluon self-energy in two non-covariant gauges, namely, CG and TAG.
The one-loop gluon self-energies calculated in CG and TAG are different 
in form from each other and have complicated expressions. However, 
we showed explicitly that once the pinch contributions are added, 
they turn out to be identical and coincide with the result previously 
obtained with covariant gauges. 
Some properties of the CG and TAG gluon self-energies were discussed 
by simply analyzing the structure of their pinch contributions. 
In the context of hot QCD, we could explain the gauge-independence of the 
hard thermal loop $\delta \Pi^{\mu \nu}$, the electric mass $m_{el}$, and 
the ``effective gluon mass'' $m_G$ from the PT point of view. 

There appear spurious singularities in the TAG gluon self-energy. 
These singularities are also present in the TAG pinch contribution. 
When the pinch contribution is added to the TAG gluon self-energy, 
the singularities cancel out. For an illustration of this cancellation, 
we calculated, in TAG, the thermal $\beta$ function in the framework 
of PT. The $\beta$ function thus obtained is indeed 
gauge-independent~\cite{Sasakia}\cite{Sasakib}. However, the result is 
incomplete in the following sense: as Elmfors and Kobes pointed out~\cite{EK}, 
the leading contribution to $\beta_T$, which gives a term $T/\kappa$, 
does not come from the hard part of the loop integral, responsible for 
a $T^2/\kappa^2$ term, but from soft loop integral. Hence it is not 
consistent to stop the calculation at one-loop order for soft internal 
momenta, and the resummed propagator and the vertices~\cite{BP} must be 
used to obtain the complete leading contribution. 
The PT algorithm still works even when we use the 
resummed propagator and the vertices~\cite{Sasakic}. 
It can be shown that the resummed effective gluon self-energy 
obtained in the framework of PT is gauge-independent and that,  
using this effective gluon self-energy,  
we can obtain the correct thermal $\beta$ function in the leading order. 
Also it can be shown that the resummed pinch contributions vanish 
on shell, and thus do not modify the result of Braaten and Pisarski~\cite{BP} 
for the gluon damping rate in the leading order.

\bigskip

\vspace{2cm}
\begin{center}
{\large\bf Acknowledgments}
\end{center}
\bigskip
\noindent
The authors would like to thank Professor A. Sirlin, 
Professor D. Zwanziger, Dr. M. Schaden and Dr. K. Philippides
for useful discussions. K.S. would like to thank Professor A. Sirlin 
for the hospitality extended to him 
at New York University where this work was done.   
This work is supported in part by Yokohama National University 
Foundation.

\newpage
\setcounter{section}{0}
\def\theequation{\Alph{section}.\arabic{equation}}
\appendix
\setcounter{equation}{0}
\section{Coulomb Gauge}
\subsection{Pinch Contribution}

\noindent
(i)The contribution of the vertices of the first kind
\bea
 \Pi^{\mu \nu (V_1)}_{P(CG)}&=&\frac{N}{2} g^2  k^2 d^{\mu \nu} \int dp 
                \biggl( \frac{-1}{p^2{\bf p}^2} + 
           \frac{-1}{q^2{\bf q}^2} \biggr)    \nonumber \\
             &+&\xi_C  \frac{N}{2} g^2  k^2 d^{\mu \nu} \int dp 
               \biggl( \frac{1}{{\bf p}^4} +
       \frac{1}{{\bf q}^4} \biggr).
\eea

\noindent
(ii)The contribution of the vertices of the second kind
\bea
  \Pi^{\mu \nu (V_2)}_{P(CG)} &=& N g^2  k^2 d^{\mu \nu} \int dp
      \frac{2}{p^2 q^2} \biggl( 
   \frac{- {\bf k}\cdot {\bf p}}{{\bf p}^2} 
  + \frac{- {\bf k}\cdot {\bf q}}{{\bf q}^2} 
              \biggr) \nonumber \\
&+& N g^2  k^2 d^{\mu \alpha} d^{\nu \beta} \int dp 
     \frac{1}{p^2 q^2 {\bf p}^2 {\bf q}^2}   
    \biggl\{ p_{\alpha}p_{\beta} 2  {\bf p}\cdot {\bf q}  
          + n_{\alpha}n_{\beta} p_0 q_0 (k^2 + 2 pq)   
   \nonumber  \\
& &\qquad + (p_{\alpha}n_{\beta}+ n_{\alpha}p_{\beta})\frac{1}{2}
  \bigl[p_0 p^2-q_0 q^2+ 2(p_0-q_0)(p_0 q_0 -  
         2 {\bf p}\cdot {\bf q} ) \bigr] \biggl\}
                \nonumber  \\
&+& \frac{N}{2} g^2 \Biggl[ d^{\mu \alpha} \int dp 
  \biggl\{ p_{\alpha}k^{\nu} \bigl[ \frac{1}{q^2  {\bf p}^2} - 
      \frac{1}{p^2  {\bf q}^2} + 
   \bigl( \frac{1}{q^2}-\frac{1}{p^2} \bigr) 
   \frac{{\bf p}\cdot {\bf q} }
   { {\bf p}^2  {\bf q}^2}  \bigr]  \nonumber \\ 
& &  + n_{\alpha}k^{\nu}\bigl[ 
  - \frac{q_0}{p^2 {\bf q}^2} -\frac{p_0}{q^2 {\bf p}^2}
     +   \bigl(\frac{q_0}{q^2}+\frac{p_0}{p^2} \bigr) 
      \frac{{\bf p}\cdot {\bf q}}
       {  {\bf p}^2 {\bf q}^2}  \bigr] \biggr\}
         + (\mu \leftrightarrow \nu) \Biggr]
          \nonumber \\
&+&\xi_C N g^2  \Biggl[
     k^2  d^{\mu \nu} \int dp 
   \biggl\{k^2 \biggl(\frac{1}{q^2 {\bf p}^4}
         +  \frac{1}{p^2 {\bf q}^4}\biggr) - 
     \frac{1}{{\bf p}^4} -  \frac{1}{{\bf q}^4} 
      \biggr\} \nonumber \\
 & &\quad  + k^2 d^{\mu \alpha} d^{\nu \beta} \int dp 
     \frac{1}{{\bf p}^2 {\bf q}^2} \Biggl\{
   p_{\alpha}p_{\beta}\biggl[k^2 \big( \frac{1}{p^2{\bf q}^2}+
     \frac{1}{q^2{\bf p}^2} \big) - \frac{1}{{\bf p}^2}-
     \frac{1}{{\bf q}^2}\biggr] 
              \nonumber  \\
& &\quad  + (p_{\alpha}n_{\beta}+ n_{\alpha}p_{\beta}) 
  \biggl[\frac{1}{2} \big(\frac{p_0}{{\bf q}^2}-
    \frac{q_0}{{\bf p}^2} \big) - k^2 \big(
   \frac{p_0}{p^2 {\bf q}^2}-\frac{q_0}{q^2 {\bf p}^2}\big)
      \biggr]    \Biggr\}   \nonumber \\
 & &\quad    + \Biggl\{ d^{\mu \alpha} \int dp 
     \frac{p_{\alpha}k^{\nu}}{2 {\bf p}^2 {\bf q}^2}  
   \bigl(\frac{{\bf k}\cdot {\bf p}}{{\bf q}^2} 
  - \frac{{\bf k}\cdot {\bf q}}{{\bf p}^2} 
     \biggr)   + (\mu \leftrightarrow \nu) 
     \Biggr\}  \Biggr]  \nonumber \\
&+&\xi_C^2 N g^2  k^4 d^{\mu \alpha} d^{\nu \beta}\int dp 
\frac{-p_{\alpha}p_{\beta}}{{\bf p}^4 {\bf q}^4}.
\eea

\noindent
(iii)The box contribution
\bea
  \Pi^{\mu \nu (Box)}_{P(CG)} &=& \frac{N}{2} g^2  k^4 d^{\mu \nu} \int dp
     \frac{1}{p^2 q^2} \biggl(\frac{1}{{\bf p}^2} 
                + \frac{1}{{\bf q}^2} \biggr) \nonumber \\   
  &+&\frac{N}{2}g^2  k^4 d^{\mu \alpha} d^{\nu \beta}\int dp 
\frac{p_{\alpha}p_{\beta} + 
   (p_{\alpha} n_{\beta}+ n_{\alpha} p_{\beta}) (q_0-p_0) 
  - 2 n_{\alpha} n_{\beta}p_0 q_0 }{p^2 q^2 {\bf p}^2 {\bf q}^2}
  \nonumber \\
 &+&\xi_C \frac{N}{2} g^2  k^4 \Biggl[ d^{\mu \nu} \int dp 
          \biggl( \frac{-1}{q^2 {\bf p}^4}+
    \frac{-1}{p^2 {\bf q}^4} \biggr)\nonumber \\
  & &+ d^{\mu \alpha} d^{\nu \beta}\int dp     
      \biggl\{ -\frac{p_{\alpha}p_{\beta}}{{\bf p}^2 {\bf q}^2} 
   \bigl(\frac{1}{p^2 {\bf q}^2}+
      \frac{1}{q^2 {\bf p}^2} \bigr) + 
\frac{p_{\alpha}n_{\beta}+n_{\alpha}p_{\beta}}
           {{\bf p}^2 {\bf q}^2} 
   \bigl(\frac{p_0}{p^2 {\bf q}^2}-
      \frac{q_0}{q^2 {\bf p}^2} \bigr)
             \biggl\}\Biggr] \nonumber \\
 &+&\xi_C^2 \frac{N}{2} g^2  k^4 d^{\mu \alpha} d^{\nu \beta}\int dp 
          \frac{p_{\alpha}p_{\beta}}{{\bf p}^4{\bf q}^4 }.
\eea        

\bigskip
\bigskip

\subsection{Expression of $\Pi^{\mu \nu}_{P(CG)}$}

The pinch contribution to the gluon self-energy in CG is rewritten in terms 
of symmetric tensors $g_{\mu\nu}$, $p_{\mu}p_{\nu}$, $q_{\mu}q_{\nu}$, 
$(p_{\mu}q_{\nu}+q_{\mu}p_{\nu})$, $(n_{\mu}p_{\nu}+p_{\mu}n_{\nu})$, 
$(n_{\mu}q_{\nu}+q_{\mu}n_{\nu})$, and $n_{\mu}n_{\nu}$.

\bea
  \Pi^{\mu \nu}_{P(CG)}(k) &=& \frac{N}{2} g^2 \int dp 
      \frac{1}{p^2 q^2 {\bf p}^2 {\bf q}^2}   \nonumber \\ 
     & & \times \Biggl[ g^{\mu\nu} k^2 \biggl\{ 
        {\bf q}^2 (k^2 - q^2 - 4 {\bf k}\cdot {\bf p}) 
           +  (p \leftrightarrow q) \biggr\}   \nonumber \\ 
  & & \quad + \biggl\{ p^{\mu}p^{\nu} \Bigl[ 
      -({\bf p}\cdot {\bf q})^2 
         -4({\bf p}\cdot {\bf q}){\bf q}^2 
            + 4 p_0q_0 q^2 + q^4 + 4 {\bf p}^2 q^2  \nonumber \\
  & &  \qquad \qquad \qquad  \qquad \qquad \qquad  
     + 3p^2 q^2 - 3 {\bf p}^2 {\bf q}^2 \Bigr]  
          +  (p \leftrightarrow q)\biggr\}   \nonumber \\
  & & \quad + ( p^{\mu}q^{\nu}+q^{\mu}p^{\nu} )
         \biggl\{ ({\bf p}\cdot {\bf q})^2 
              - 2 ({\bf p}\cdot {\bf q}) 
                      ({\bf p}^2 + {\bf q}^2) 
       - 5 {\bf p}^2 {\bf q}^2  \biggr\} \nonumber \\
  & & \quad + \biggl\{ (n^{\mu}p^{\nu}+p^{\mu}n^{\nu}) 
       \Bigl[ -kq \Bigl(p^2 q_0 - q^2 p_0 
            -2 {\bf p}\cdot {\bf q} (p_0-q_0) \Bigr) 
    +4k_0(pq)p_0q_0   \nonumber \\
  & & \qquad \qquad \qquad + q^2{\bf p}^2q_0 
       + p^2{\bf q}^2p_0 -(p^2 q_0 + q^2 p_0) 
             {\bf p}\cdot {\bf q} \Bigr]
     + (p \leftrightarrow q)  \biggr\}  \nonumber \\
  & & \quad + n^{\mu}n^{\nu} 4k^2(pq)p_0q_0 \Biggr]   \nonumber \\
 &+&\xi_C \frac{N}{2} g^2  \int dp 
     \frac{1}{p^2 q^2 {\bf p}^4 {\bf q}^4}   \nonumber \\ 
     & & \times \Biggl[ g^{\mu\nu} k^2 \biggl\{ 
             k^2(p^2 {\bf q}^4 + q^2  {\bf p}^4) 
                 - p^2 q^2 ({\bf p}^4 + {\bf q}^4 )
                              \biggr\}   \nonumber \\ 
  & & \quad + \biggl\{ p^{\mu}p^{\nu}  \Bigl[
   q^2 {\bf p}^2 \Bigl(2kq(k_0p_0)+(kq)^2-k^2{\bf p}^2 \Bigr)
  \nonumber \\
    & &  \qquad \qquad \qquad  \qquad
 + p^2 {\bf p}^2 \Bigl(-2kq(k_0q_0)+(kq)^2-k^2{\bf q}^2 \Bigr)
                                 \nonumber \\
  & &  \qquad \qquad \qquad  \qquad  
     + p^2q^2 \Bigl( {\bf q}^4+ 
               {\bf p}^2 (2kq+{\bf p}^2) \Bigr)  \Bigr] 
          +  (p \leftrightarrow q) \biggr\}   \nonumber \\
  & & \quad + ( p^{\mu}q^{\nu}+q^{\mu}p^{\nu} )
         \biggl\{ p^2 {\bf q}^2 
            \Bigl[(k_0q_0)(q^2-p^2)-(kp)(kq)-k^2{\bf q}^2\Bigr]
     \nonumber \\
   & &  \qquad \qquad \qquad  \qquad \qquad \qquad
               +p^2q^2{\bf q}^2\Bigl[{\bf q}^2-kq\Bigr] 
                     + (p \leftrightarrow q) \biggr\} \nonumber \\
  & & \quad + \biggl\{ (n^{\mu}p^{\nu}+p^{\mu}n^{\nu}) 
        kq  \Bigl[ k^2(q^2 {\bf p}^2 p_0 
                               - p^2 {\bf q}^2 q_0 )
      \nonumber  \\
  & &  \qquad \qquad \qquad  \qquad \qquad \qquad
      -p^2q^2({\bf p}^2 p_0-{\bf q}^2 q_0 )  \Bigr] 
     + (p \leftrightarrow q) \biggr\}  \Biggr] \nonumber \\
&+&\xi_C^2 \frac{N}{2} g^2 \int dp 
      \frac{1}{{\bf p}^4 {\bf q}^4} \nonumber \\
          & & \times \Biggl[ \biggl\{-(kq)^2  p^{\mu}p^{\nu} 
              + (p \leftrightarrow q) \biggr\} 
        + (kp)(kq) (p^{\mu}q^{\nu}+q^{\mu}p^{\nu}) \Biggr].
\eea

\setcounter{equation}{0}
\section{Temporal Axial Gauge}

\subsection{Pinch contribution}

Note that the gauge parameter $\xi_A$ has a dimension ${\rm mass}^{-2}$.

\noindent
(i)The pinch contribution from the vertices of the first kind
\bea
 \Pi^{\mu \nu (V_1)}_{P(TAG)}&=&\frac{N}{2} g^2  k^2 d^{\mu \nu} \int dp 
   \biggl(\frac{-1}{p^2 p_0^2} + \frac{-1}{q^2 q_0^2} \biggr)\nonumber \\
             &+&\xi_A \frac{N}{2} g^2  k^2 d^{\mu \nu} \int dp 
   \biggl(\frac{-1}{p_0^2} + \frac{-1}{q_0^2}  \biggr).
\eea

\noindent
(ii)The contribution of the vertices of the second kind
\bea
  \Pi^{\mu \nu (V_2)}_{P(TAG)} &=& N g^2  k^2 d^{\mu \nu} \int dp
      \frac{1}{p^2 q^2} \biggl\{ \frac{p^2}{p_0^2}+ \frac{q^2}{q_0^2}
  - \frac{2{\bf k}\cdot {\bf p}}{p_0^2} 
  - \frac{2 {\bf k}\cdot {\bf q}}{q_0^2} 
              \biggr\} \nonumber \\
&+& N g^2  k^2 d^{\mu \alpha} d^{\nu \beta} \int dp 
     \frac{1}{p^2 q^2 p_0^2 q_0^2}   
    \biggl\{ p_{\alpha}p_{\beta}   \bigl[ 
         {\bf k}^2 - p_0^2 - q_0^2 \bigr] + 
   n_{\alpha}n_{\beta} p_0 q_0 (k^2 + 2 pq)   
   \nonumber  \\
& &\quad + (p_{\alpha}n_{\beta}+ n_{\alpha}p_{\beta})
  \bigl[ \frac{1}{2} (p_0 p^2-q_0 q^2) -p_0 ({\bf q}^2+
  2{\bf p}\cdot {\bf q}) 
   +q_0 ({\bf p}^2+ 2{\bf p}\cdot {\bf q})               
                \bigr] \biggl\}
                \nonumber  \\
 &+&\xi_A N g^2  \Biggl[
     k^2  d^{\mu \nu} \int dp 
   \biggl\{-k^2 \biggl(\frac{1}{q^2 p_0^2}
         +  \frac{1}{p^2 q_0^2}\biggr) +
     \frac{1}{p_0^2} + \frac{1}{q_0^2} 
      \biggr\} \nonumber \\
 & &\quad  + k^2 d^{\mu \alpha} d^{\nu \beta} \int dp 
     \frac{1}{p_0^2 q_0^2} \Biggl\{
   p_{\alpha}p_{\beta}\biggl[-k^2 \bigl( \frac{1}{p^2}+
     \frac{1}{q^2} \bigr) \biggr] 
              \nonumber  \\
& &\quad  + (p_{\alpha}n_{\beta}+ n_{\alpha}p_{\beta}) 
  \biggl[\frac{1}{2}(p_0-q_0) +p_0 \frac{q^2+2pq}{p^2}-
      q_0\frac{p^2+2pq}{q^2}
                 \biggr]    \Biggr\}   \Biggr]  \nonumber \\
&+&\xi_A^2 N g^2  k^4 d^{\mu \alpha} d^{\nu \beta}\int dp 
\frac{-p_{\alpha}p_{\beta}}{p_0^2 q_0^2}.
\eea

\noindent
(iii) The pinch contribution from the box diagrams
\bea
  \Pi^{\mu \nu (Box)}_{P(TAG)} &=& \frac{N}{2} g^2  k^4 d^{\mu \nu} \int dp
     \frac{1}{p^2 q^2}\biggl(\frac{1}{p_0^2}+\frac{1}{q_0^2}\biggr)  
                     \nonumber \\   
  &+&\frac{N}{2}g^2  k^4 d^{\mu \alpha} d^{\nu \beta}\int dp 
\frac{p_{\alpha}p_{\beta} + 
   (p_{\alpha} n_{\beta}+ n_{\alpha} p_{\beta}) (q_0-p_0) 
  - 2 n_{\alpha} n_{\beta}p_0 q_0 }{p^2 q^2 p_0^2 q_0^2}
  \nonumber \\
 &+&\xi_A \frac{N}{2} g^2  k^4 \Biggl[ d^{\mu \nu} \int dp 
      \biggl(\frac{1}{q^2 p_0^2}+ \frac{1}{p^2 q_0^2} \biggr) \nonumber \\
  & &+ d^{\mu \alpha} d^{\nu \beta}\int dp     
       \biggl\{ \frac{p_{\alpha}p_{\beta}}{p_0^2 q_0^2} 
        \bigl(\frac{1}{p^2}+\frac{1}{q^2}\bigr) + 
    \frac{ p_{\alpha} n_{\beta}+ n_{\alpha} p_{\beta}}
        {p_0^2 q_0^2 }\bigl(\frac{q_0}{q^2}-\frac{p_0}{p^2}\bigr)
            \biggr\} \Biggr] \nonumber \\
 &+&\xi_A^2 \frac{N}{2} g^2  k^4 d^{\mu \alpha} d^{\nu \beta}\int dp 
          \frac{p_{\alpha}p_{\beta}}{p_0^2 q_0^2}.
\eea

\subsection{Expression of $\Pi^{\mu \nu}_{P(TAG)}$}

The pinch contribution to the gluon self-energy in TAG 
is rewritten in terms of symmetric tensors 
$g_{\mu\nu}$, $p_{\mu}p_{\nu}$, $q_{\mu}q_{\nu}$, 
$(p_{\mu}q_{\nu}+q_{\mu}p_{\nu})$, $(n_{\mu}p_{\nu}+p_{\mu}n_{\nu})$, 
$(n_{\mu}q_{\nu}+q_{\mu}n_{\nu})$, and $n_{\mu}n_{\nu}$.

\bea
  \Pi^{\mu \nu}_{P(TAG)}(k) &=& \frac{N}{2} g^2 \int dp 
                 \frac{1}{p^2 q^2}   \nonumber \\
  & & \times \Biggl[ g^{\mu\nu} \biggl\{ 
            \frac{k^2 (k^2 +2p^2 - q^2 -
     4 {\bf k}\cdot {\bf p}}{p_0^2}  
         +   (p \leftrightarrow q)   \biggr\} \nonumber \\ 
  & & \quad + \biggl\{ p^{\mu}p^{\nu} \biggl[ -3 -
     \frac{({\bf p}\cdot {\bf q})^2 }{p_0^2 q_0^2}
   +\frac{ 2{\bf p}\cdot {\bf q}}{p_0^2 q_0^2}
         (q^2-2q_0^2) -\frac{q^4}{p_0^2 q_0^2}    \nonumber \\ 
 & &  \qquad \qquad \qquad \qquad \qquad \qquad \qquad \qquad
  +\frac{p^2}{p_0^2}-\frac{q^2}{q_0^2}
 \biggr] + (p \leftrightarrow q)  \biggr\}  \nonumber \\
 & & \quad + ( p^{\mu}q^{\nu}+q^{\mu}p^{\nu} ) 
   \biggl[ -5+ \frac{({\bf p}\cdot {\bf q})^2 }
             {p_0^2 q_0^2} 
     -({\bf p}\cdot {\bf q}) \Bigl(
        \frac{p^2+q^2}{p_0^2 q_0^2}+ 
           \frac{2}{p_0^2} + \frac{2}{q_0^2} \Bigr)
   +\frac{p^2q^2}{p_0^2 q_0^2} \biggr]  \nonumber  \\
 & & \quad + \biggl\{ (n^{\mu}p^{\nu}+p^{\mu}n^{\nu}) 
         \frac{1}{p_0^2 q_0^2} 
     \biggl[ -kq \Bigl( -p^2 q_0 + q^2 p_0 
            -2 {\bf p}\cdot {\bf q} (p_0-q_0) \Bigr) 
   \nonumber \\
  & & \qquad \qquad \qquad \qquad \qquad \qquad \qquad \quad +4k_0 p_0q_0 (pq) 
     \biggr] +  (p \leftrightarrow q)  \biggr\}  \nonumber \\
 & & \quad + n^{\mu}n^{\nu} \frac{4p_0q_0k^2(pq)}
   {p_0^2 q_0^2}  \qquad  \Biggr]   
    \nonumber \\
  &+&\xi_A \frac{N}{2} g^2  \int dp 
      \Biggl[ g^{\mu\nu} \biggl\{ \biggl(-k^4 \frac{1}{q^2 p_0^2}
    +k^2\frac{1}{p_0^2}
            \biggr) + (p \leftrightarrow q)  \biggr\}  \nonumber \\
& & \quad + \biggl\{ p^{\mu}p^{\nu}  \biggl[ 
    +k^2 \Bigl(\frac{1}{p^2q_0^2}+ \frac{1}{q^2p_0^2}  \Bigr)
    -\frac{(kq)^2}{p_0^2 q_0^2}\Bigl(\frac{1}{p^2}+ \frac{1}{q^2} \Bigr)
         \nonumber  \\
 & &  \qquad  \qquad \qquad \qquad \qquad + \frac{2k_0(kq)}{p_0^2 q_0^2}
     \Bigl( \frac{q_0}{q^2}-\frac{p_0}{p^2} \Bigr) 
    -\frac{3}{p_0^2}-\frac{1}{q_0^2}
      \biggr] + (p \leftrightarrow q)  \biggr\}  \nonumber \\
   & & \quad + ( p^{\mu}q^{\nu}+q^{\mu}p^{\nu} )
    \biggl[\biggl(\frac{+k^2p_0^2+(kp)(kq)-k_0p_0(p^2-q^2)}{p^2p_0^2q_0^2}
      -\frac{2}{p_0^2}
    \biggr) + (p \leftrightarrow q)  \biggr] \nonumber \\
 & & \quad + \biggl\{ (n^{\mu}p^{\nu}+p^{\mu}n^{\nu})
         \frac{kq}{p_0^2q_0^2}
   \biggl[-k^2  \Bigl( \frac{p_0}{p^2} - \frac{q_0}{q^2}    \Bigr)
 -q_0+p_0 \biggr] + (p \leftrightarrow q)  \biggr\} \Biggr]  \nonumber \\ 
 &+&\xi_A^2 \frac{N}{2} g^2 \int dp 
      \frac{1}{p_0^2 q_0^2} 
           \biggl[ \biggl\{-(kq)^2  p^{\mu}p^{\nu} 
              + (p \leftrightarrow q) \biggr\} 
        +(kp)(kq) (p^{\mu}q^{\nu}+q^{\mu}p^{\nu}) \biggr]~. \nonumber \\
\eea

\setcounter{equation}{0}
\section{Thermal one-loop integrals}

We list the thermal one-loop integrals 
in the static limit $k_0=0$ which appear in Sec.5. 
The expressions are in the imaginary time formalism and thus
\be 
     \int dp =\int \frac{d^3 p}{(2\pi)^3} ~T \sum_{n},
\ee 
where the summation goes over $p_0=2\pi inT$. 
We only give the matter part.
Due to the constraint $k+p+q=0$ we have
\be
       \int dp f(p,q) = \int dp f(q,p).
\ee
It is understood that 
in the r.h.s. of the expressions below, $p\equiv \vert  {\bf p} \vert$,  
$\kappa\equiv \vert  {\bf k} \vert$ and $n(p)=1/[{\rm exp} (p/T) -1 ]$~.

\bea
    {\bf k}^2 \int dp \frac{ {\bf k}^2 + 
  4  {\bf k} \cdot  {\bf p}}{p^2q^2p_0^2} &=&
   \frac{1}{4\pi^2}\int_{0}^{\infty} dp~p~n(p)
  \biggl( -\frac{\kappa^3}{p^3} \biggr)
    ~{\rm ln} \Big\vert \frac{2p + \kappa}{2p - \kappa} \Big\vert  \\
\  \nonumber \\
    {\bf k}^2 \int dp \frac{1}{q^2p_0^2} &=&
   \frac{1}{4\pi^2}\int_{0}^{\infty} dp~p~n(p)
      \biggl( -2\frac{\kappa^2}{p^2} \biggr)  \\
\ \nonumber \\
    {\bf k}^2 \int dp \frac{1}{p^2p_0^2} &=&
   \frac{1}{4\pi^2}\int_{0}^{\infty} dp~p~n(p)
      \biggl( -2\frac{\kappa^2}{p^2} \biggr)  
\eea
\bea
& &  {\bf k}^4 \int dp \biggl[
         {\bf p}^2-\frac{( {\bf k} \cdot  {\bf p})^2}
        { {\bf k}^2} \biggr] \frac{1}{p^2q^2p_0^2q_0^2} 
    \nonumber \\
& & \qquad \qquad  = \frac{1}{4\pi^2}\int_{0}^{\infty} dp~p~n(p)
   \Biggl\{\frac{\kappa^4}{p^4} +\frac{\kappa^3(4p^2-\kappa^2)}{4p^5}
   ~{\rm ln} \Big\vert \frac{2p + \kappa}{2p - \kappa} \Big\vert 
    \Biggr\}  \\
\  \nonumber \\
& &  {\bf k}^2 \int dp \biggl[
         {\bf p}^2-\frac{( {\bf k} \cdot  {\bf p})^2}
        { {\bf k}^2} \biggr] \frac{1}{p^2q^2p_0^2} 
  \nonumber \\
& & \qquad \qquad  = \frac{1}{4\pi^2}\int_{0}^{\infty} dp~p~n(p)
    \Biggl\{\frac{\kappa^2}{p^2} +\frac{\kappa(4p^2-\kappa^2)}{4p^3}
   ~{\rm ln} \Big\vert \frac{2p + \kappa}{2p - \kappa} \Big\vert 
    \Biggr\}.
\eea


\newpage
\noindent
{\large\bf Figure Caption}
\medskip

\noindent
Fig.1

\noindent
(a) The gluon self-energy diagrams for the quark-quark scattering. 
(b) The gluon self-energy diagram with three-gluon interactions. 
(c) The tadpole diagram for the gluon self-energy. 
(d) The ghost diagram for the gluon self-energy. 

\medskip

\noindent
Fig.2

\noindent
(a) The vertex diagrams of the first kind for 
the quark-quark scattering. 
(b) Their pinch contribution.

\medskip

\noindent
Fig.3

\noindent
(a) The vertex diagram of the second kind 
for the quark-quark scattering. 
(b) Its pinch contribution.

\medskip

\noindent
Fig.4

\noindent
(a) The box diagrams for the quark-quark scattering.   
(b) Their pinch contribution.

\medskip

\noindent
Fig.5

\noindent
(a) The ghost-gluon vertex in Coulomb gauge; (b) in temporal
axial gauge.



\newenvironment{texdraw}{\leavevmode\btexdraw}{\etexdraw}

\def\setRevDate $#1 #2 #3${\def\TeXdrawId{TeXdraw V1R3 revised <#2>}}
\setRevDate $Date: Wed Mar  3 12:01:12 MST 1993$

\chardef\catamp=\the\catcode`\@
\catcode`\@=11
\ifx\TeXdraw@included\undefined\global\let\TeXdraw@included=\relax\else
\errhelp{TeXdraw needs to be input only once outside of any groups.}%
\errmessage{Multiple call to include TeXdraw ignored}%
\expandafter\endinput\fi
\long
\def\centertexdraw #1{\hbox to \hsize{\hss
\btexdraw #1\etexdraw
\hss}}
\def\btexdraw {\x@pix=0 \y@pix=0
\x@segoffpix=\x@pix  \y@segoffpix=\y@pix
\t@exdrawdef
\setbox\t@xdbox=\vbox\bgroup\offinterlineskip
\global\d@bs=0 
\t@extonlytrue 
\p@osinitfalse
\savemove \x@pix \y@pix 
\m@pendingfalse
\p@osinitfalse          
\p@athfalse}
\def\etexdraw {\ift@extonly \else
\t@drclose
\fi
\egroup 
\ifdim \wd\t@xdbox>0pt
\errmessage{TeXdraw box non-zero size, possible extraneous text}%
\fi
\maxhvpos 
\pixtodim \xminpix \l@lxpos  \pixtodim \yminpix \l@lypos
\pixtobp {-\xminpix}\l@lxbp  \pixtobp {-\yminpix}\l@lybp
\vbox {\offinterlineskip
\ift@extonly \else
\includepsfile{\p@sfile}{\the\l@lxbp}{\the\l@lybp}%
{\the\hdrawsize}{\the\vdrawsize}%
\fi
\vskip\vdrawsize
\vskip \l@lypos
\hbox {\hskip -\l@lxpos
\box\t@xdbox
\hskip \hdrawsize
\hskip \l@lxpos}%
\vskip -\l@lypos\relax}}
\def\drawdim #1 {\def\d@dim{#1\relax}}
\def\setunitscale #1 {\edef\u@nitsc{#1}%
\realmult \u@nitsc  \s@egsc \d@sc}
\def\relunitscale #1 {\realmult {#1}\u@nitsc \u@nitsc
\realmult \u@nitsc \s@egsc \d@sc}
\def\setsegscale #1 {\edef\s@egsc {#1}%
\realmult \u@nitsc \s@egsc \d@sc}
\def\relsegscale #1 {\realmult {#1}\s@egsc \s@egsc
\realmult \u@nitsc \s@egsc \d@sc}
\def\bsegment {\ifp@ath
\flushmove
\fi
\begingroup
\x@segoffpix=\x@pix
\y@segoffpix=\y@pix
\setsegscale 1
\global\advance \d@bs by 1 }
\def\esegment {\endgroup
\ifnum \d@bs=0
\writetx {es}%
\else
\global\advance \d@bs by -1
\fi}
\def\savecurrpos (#1 #2){\getsympos (#1 #2)\a@rgx\a@rgy
\s@etcsn \a@rgx {\the\x@pix}%
\s@etcsn \a@rgy {\the\y@pix}}%
\def\savepos (#1 #2)(#3 #4){\getpos (#1 #2)\a@rgx\a@rgy
\coordtopix \a@rgx \t@pixa
\advance \t@pixa by \x@segoffpix
\coordtopix \a@rgy \t@pixb
\advance \t@pixb by \y@segoffpix
\getsympos (#3 #4)\a@rgx\a@rgy
\s@etcsn \a@rgx {\the\t@pixa}%
\s@etcsn \a@rgy {\the\t@pixb}}
\def\linewd #1 {\coordtopix {#1}\t@pixa
\flushbs
\writetx {\the\t@pixa\space sl}}
\def\setgray #1 {\flushbs
\writetx {#1 sg}}
\def\lpatt (#1){\listtopix (#1)\p@ixlist
\flushbs
\writetx {[\p@ixlist] sd}}
\def\lvec (#1 #2){\getpos (#1 #2)\a@rgx\a@rgy
\s@etpospix \a@rgx \a@rgy
\writeps {\the\x@pix\space \the\y@pix\space lv}}
\def\rlvec (#1 #2){\getpos (#1 #2)\a@rgx\a@rgy
\r@elpospix \a@rgx \a@rgy
\writeps {\the\x@pix\space \the\y@pix\space lv}}
\def\move (#1 #2){\getpos (#1 #2)\a@rgx\a@rgy
\s@etpospix \a@rgx \a@rgy
\savemove \x@pix \y@pix}
\def\rmove (#1 #2){\getpos (#1 #2)\a@rgx\a@rgy
\r@elpospix \a@rgx \a@rgy
\savemove \x@pix \y@pix}
\def\lcir r:#1 {\coordtopix {#1}\t@pixa
\writeps {\the\t@pixa\space cr}%
\r@elupd \t@pixa \t@pixa
\r@elupd {-\t@pixa}{-\t@pixa}}
\def\fcir f:#1 r:#2 {\coordtopix {#2}\t@pixa
\writeps {#1 \the\t@pixa\space fc}%
\r@elupd \t@pixa \t@pixa
\r@elupd {-\t@pixa}{-\t@pixa}}
\def\lellip rx:#1 ry:#2 {\coordtopix {#1}\t@pixa
\coordtopix {#2}\t@pixb
\writeps {\the\t@pixa\space \the\t@pixb\space el}%
\r@elupd \t@pixa \t@pixb
\r@elupd {-\t@pixa}{-\t@pixb}}
\def\larc r:#1 sd:#2 ed:#3 {\coordtopix {#1}\t@pixa
\writeps {\the\t@pixa\space #2 #3 ar}}
\def\ifill f:#1 {\writeps {#1 fl}}
\def\lfill f:#1 {\writeps {#1 fp}}
\def\htext #1{\def\testit {#1}%
\ifx \testit\l@paren
\let\next=\h@move
\else
\let\next=\h@text
\fi
\next{#1}}
\def\rtext td:#1 #2{\def\testit {#2}%
\ifx \testit\l@paren
\let\next=\r@move
\else
\let\next=\r@text
\fi
\next td:#1 {#2}}
\def\vtext {\rtext td:90 }
\def\textref h:#1 v:#2 {\ifx #1R%
\edef\l@stuff {\hss}\edef\r@stuff {}%
\else
\ifx #1C%
\edef\l@stuff {\hss}\edef\r@stuff {\hss}%
\else  
\edef\l@stuff {}\edef\r@stuff {\hss}%
\fi
\fi
\ifx #2T%
\edef\t@stuff {}\edef\b@stuff {\vss}%
\else
\ifx #2C%
\edef\t@stuff {\vss}\edef\b@stuff {\vss}%
\else  
\edef\t@stuff {\vss}\edef\b@stuff {}%
\fi
\fi}
\def\avec (#1 #2){\getpos (#1 #2)\a@rgx\a@rgy
\s@etpospix \a@rgx \a@rgy
\writeps {\the\x@pix\space \the\y@pix\space (\a@type)
\the\a@lenpix\space \the\a@widpix\space av}}
\def\ravec (#1 #2){\getpos (#1 #2)\a@rgx\a@rgy
\r@elpospix \a@rgx \a@rgy
\writeps {\the\x@pix\space \the\y@pix\space (\a@type)
\the\a@lenpix\space \the\a@widpix\space av}}
\def\arrowheadsize l:#1 w:#2 {\coordtopix{#1}\a@lenpix
\coordtopix{#2}\a@widpix}
\def\arrowheadtype t:#1 {\edef\a@type{#1}}
\def\curvytype#1{\def\curv@type{#1}}\curvytype{4}%
\def\curvyheight#1{\def\curv@height{#1}}\curvyheight{10}%
\def\curvylength#1{\def\curv@length{#1}}\curvylength{10}%
\def\drawcurvyphoton around (#1 #2) from (#3 #4) to (#5 #6)%
{\getpos (#1 #2)\a@rgx\a@rgy
\coordtopix \a@rgx \t@pixa \advance \t@pixa by \x@segoffpix
\coordtopix \a@rgy \t@pixb \advance \t@pixb by \y@segoffpix
\writeps {mark \the\t@pixa\space \the\t@pixb}%
\getpos (#3 #4)\a@rgx\a@rgy
\s@etpospix \a@rgx\a@rgy
\writeps {\the\x@pix\space \the\y@pix}%
\getpos (#5 #6)\a@rgx\a@rgy
\s@etpospix \a@rgx\a@rgy
\writeps {\the\x@pix\space \the\y@pix}%
\writeps {\curv@height\space \curv@length\space \curv@type\space%
curvyphoton}%
}%
\def\drawcurvygluon around (#1 #2) from (#3 #4) to (#5 #6)%
{\getpos (#1 #2)\a@rgx\a@rgy
\coordtopix \a@rgx \t@pixa \advance \t@pixa by \x@segoffpix
\coordtopix \a@rgy \t@pixb \advance \t@pixb by \y@segoffpix
\writeps {mark \the\t@pixa\space \the\t@pixb}%
\getpos (#3 #4)\a@rgx\a@rgy
\s@etpospix \a@rgx\a@rgy
\writeps {\the\x@pix\space \the\y@pix}%
\getpos (#5 #6)\a@rgx\a@rgy
\s@etpospix \a@rgx\a@rgy
\writeps {\the\x@pix\space \the\y@pix}%
\writeps {\curv@height\space 2 mul \curv@length\space \curv@type\space%
curvygluon}%
}%
\def\blobfreq#1{\def\bl@bfreq{#1}}\blobfreq{0.2}%
\def\blobangle#1{\def\bl@bangle{#1}}\blobangle{0}%
\def\hatchedblob#1{\def\bl@btype{(#1)}}\hatchedblob{B}%
\def\grayblob#1{\def\bl@btype{#1}}%
\def\drawblob xsize:#1 ysize:#2 at (#3 #4)%
{\getpos (#3 #4)\a@rgx\a@rgy \s@etpospix \a@rgx\a@rgy
\writeps{\the\x@pix\space \the\y@pix}%
\getpos (#1 #2)\a@rgx\a@rgy 
\coordtopix \a@rgx\t@pixa \coordtopix\a@rgy\t@pixb \writeps
{\the\t@pixa\space\the\t@pixb\space\bl@bangle\space\bl@bfreq\space\bl@btype}%
\writeps {blob}%
\rmove (-#1 -#2)\rmove (#1 #2)\rmove (#1 #2)\rmove (-#1 -#2)}%
\def\drawbb {\bsegment
\drawdim bp
\setunitscale 0.24
\linewd 1
\writeps {\the\xminpix\space \the\yminpix\space mv}%
\writeps {\the\xminpix\space \the\ymaxpix\space lv}%
\writeps {\the\xmaxpix\space \the\ymaxpix\space lv}%
\writeps {\the\xmaxpix\space \the\yminpix\space lv}%
\writeps {\the\xminpix\space \the\yminpix\space lv}%
\esegment}
\def\getpos (#1 #2)#3#4{\g@etargxy #1 #2 {} \\#3#4%
\c@heckast #3%
\ifa@st
\g@etsympix #3\t@pixa
\advance \t@pixa by -\x@segoffpix
\pixtocoord \t@pixa #3
\fi
\c@heckast #4%
\ifa@st
\g@etsympix #4\t@pixa
\advance \t@pixa by -\y@segoffpix
\pixtocoord \t@pixa #4
\fi}
\def\getsympos (#1 #2)#3#4{\g@etargxy #1 #2 {} \\#3#4%
\c@heckast #3%
\ifa@st \else
\errmessage {TeXdraw: invalid symbolic coordinate}
\fi
\c@heckast #4%
\ifa@st \else
\errmessage {TeXdraw: invalid symbolic coordinate}
\fi}
\def\listtopix (#1)#2{\def #2{}%
\edef\l@ist {#1 }%
\t@countc=0
\loop
\expandafter\g@etitem \l@ist \\\a@rgx\l@ist
\a@pppix \a@rgx #2
\ifx \l@ist\empty
\t@countc=1
\fi
\ifnum \t@countc=0
\repeat}
\def\realmult #1#2#3{\dimen0=#1pt
\dimen2=#2\dimen0
\edef #3{\expandafter\c@lean\the\dimen2}}
\def\intdiv #1#2#3{\t@counta=#1
\t@countb=#2
\ifnum \t@countb<0
\t@counta=-\t@counta
\t@countb=-\t@countb
\fi
\t@countd=1
\ifnum \t@counta<0
\t@counta=-\t@counta
\t@countd=-1
\fi
\t@countc=\t@counta  \divide \t@countc by \t@countb
\t@counte=\t@countc  \multiply \t@counte by \t@countb
\advance \t@counta by -\t@counte
\t@counte=-1
\loop
\advance \t@counte by 1
\ifnum \t@counte<16
\multiply \t@countc by 2 
\multiply \t@counta by 2 
\ifnum \t@counta<\t@countb \else
\advance \t@countc by 1     
\advance \t@counta by -\t@countb 
\fi
\repeat
\divide \t@countb by 2        
\ifnum \t@counta<\t@countb    
\advance \t@countc by 1
\fi
\ifnum \t@countd<0            
\t@countc=-\t@countc
\fi
\dimen0=\t@countc sp          
\edef #3{\expandafter\c@lean\the\dimen0}}
\outer\def\gnewif #1{\count@\escapechar \escapechar\m@ne
\expandafter\expandafter\expandafter
\edef\@if #1{true}{\global\let\noexpand#1=\noexpand\iftrue}%
\expandafter\expandafter\expandafter
\edef\@if #1{false}{\global\let\noexpand#1=\noexpand\iffalse}%
\@if#1{false}\escapechar\count@} 
\def\@if #1#2{\csname\expandafter\if@\string#1#2\endcsname}
{\uccode`1=`i \uccode`2=`f \uppercase{\gdef\if@12{}}} 
\def\coordtopix #1#2{\dimen0=#1\d@dim
\dimen2=\d@sc\dimen0
\t@counta=\dimen2    
\t@countb=\s@ppix
\divide \t@countb by 2
\ifnum \t@counta<0  
\advance \t@counta by -\t@countb
\else
\advance \t@counta by \t@countb
\fi
\divide \t@counta by \s@ppix
#2=\t@counta}
\def\pixtocoord #1#2{\t@counta=#1%
\multiply \t@counta by \s@ppix
\dimen0=\d@sc\d@dim
\t@countb=\dimen0
\intdiv \t@counta \t@countb #2}
\def\pixtodim #1#2{\t@countb=#1%
\multiply \t@countb by \s@ppix
#2=\t@countb sp\relax}
\def\pixtobp #1#2{\dimen0=\p@sfactor pt
\t@counta=\dimen0
\multiply \t@counta by #1%
\ifnum \t@counta < 0
\advance \t@counta by -32768
\else
\advance \t@counta by 32768
\fi
\divide \t@counta by 65536
#2=\t@counta}
\newcount\t@counta \newcount\t@countb
\newcount\t@countc \newcount\t@countd
\newcount\t@counte
\newcount\t@pixa \newcount\t@pixb
\newcount\t@pixc \newcount\t@pixd
\let\l@lxbp=\t@pixa \let\l@lybp=\t@pixb 
\let\u@rxbp=\t@pixc \let\u@rybp=\t@pixd
\newdimen\t@xpos \newdimen\t@ypos
\let\l@lxpos=\t@xpos \let\l@lypos=\t@ypos
\newcount\xminpix \newcount\xmaxpix
\newcount\yminpix \newcount\ymaxpix
\newcount\a@lenpix \newcount\a@widpix
\newcount\x@pix \newcount\y@pix
\newcount\x@segoffpix \newcount\y@segoffpix
\newcount\x@savepix \newcount\y@savepix
\newcount\s@ppix 
\newcount\d@bs
\newcount\t@xdnum
\global\t@xdnum=0
\newdimen\hdrawsize \newdimen\vdrawsize
\newbox\t@xdbox
\newwrite\drawfile
\newif\ifm@pending
\newif\ifp@ath
\newif\ifa@st
\gnewif \ift@extonly
\gnewif\ifp@osinit
\def\l@paren{(}
\def\a@st{*}
\catcode`\%=12
  \def\p@b {
\catcode`\%=14
\catcode`\{=12 \catcode`\}=12  \catcode`\u=1 \catcode`\v=2
  \def\l@br u{v \def\r@br u}v
\catcode `\{=1 \catcode`\}=2   \catcode`\u=11 \catcode`\v=11
{\catcode`\p=12 \catcode`\t=12
 \gdef\c@lean #1pt{#1}}
\def\sppix#1/#2 {\dimen0=1#2 \s@ppix=\dimen0
\t@counta=#1%
\divide \t@counta by 2
\advance \s@ppix by \t@counta
\divide \s@ppix by #1%
\t@counta=\s@ppix
\multiply \t@counta by 65536 
\advance \t@counta by 32891  
\divide \t@counta by 65782 
\dimen0=\t@counta sp
\edef\p@sfactor {\expandafter\c@lean\the\dimen0}}
\def\g@etargxy #1 #2 #3 #4\\#5#6{\def #5{#1}%
\ifx #5\empty
\g@etargxy #2 #3 #4 \\#5#6%
\else
\def #6{#2}%
\def\next {#3}%
\ifx \next\empty \else
\errmessage {TeXdraw: invalid coordinate}%
\fi
\fi}
\def\c@heckast #1{\expandafter
\c@heckastll #1\\}
\def\c@heckastll #1#2\\{\def\testit {#1}%
\ifx \testit\a@st
\a@sttrue
\else
\a@stfalse
\fi}
\def\g@etsympix #1#2{\expandafter
\ifx \csname #1\endcsname \relax
\errmessage {TeXdraw: undefined symbolic coordinate}%
\fi
#2=\csname #1\endcsname}
\def\s@etcsn #1#2{\expandafter
\xdef\csname#1\endcsname {#2}}
\def\g@etitem #1 #2\\#3#4{\edef #4{#2}\edef #3{#1}}
\def\a@pppix #1#2{\edef\next  {#1}%
\ifx \next\empty \else
\coordtopix {#1}\t@pixa
\ifx #2\empty
\edef #2{\the\t@pixa}%
\else
\edef #2{#2 \the\t@pixa}%
\fi
\fi}
\def\s@etpospix #1#2{\coordtopix {#1}\x@pix
\advance \x@pix by \x@segoffpix
\coordtopix {#2}\y@pix
\advance \y@pix by \y@segoffpix
\u@pdateminmax \x@pix \y@pix}
\def\r@elpospix #1#2{\coordtopix {#1}\t@pixa
\advance \x@pix by \t@pixa
\coordtopix {#2}\t@pixa
\advance \y@pix by \t@pixa
\u@pdateminmax \x@pix \y@pix}
\def\r@elupd #1#2{\t@counta=\x@pix
\advance\t@counta by #1%
\t@countb=\y@pix
\advance\t@countb by #2%
\u@pdateminmax \t@counta \t@countb}
\def\u@pdateminmax #1#2{\ifnum #1>\xmaxpix
\global\xmaxpix=#1%
\fi
\ifnum #1<\xminpix
\global\xminpix=#1%
\fi
\ifnum #2>\ymaxpix
\global\ymaxpix=#2%
\fi
\ifnum #2<\yminpix
\global\yminpix=#2%
\fi}
\def\maxhvpos {\t@pixa=\xmaxpix
\advance \t@pixa by -\xminpix
\pixtodim  \t@pixa {\dimen2}%
\global\hdrawsize=\dimen2
\t@pixa=\ymaxpix
\advance \t@pixa by -\yminpix
\pixtodim \t@pixa {\dimen2}%
\global\vdrawsize=\dimen2\relax}
\def\savemove #1#2{\x@savepix=#1\y@savepix=#2%
\m@pendingtrue
\ifp@osinit \else
\p@osinittrue
\global\xminpix=\x@savepix \global\yminpix=\y@savepix
\global\xmaxpix=\x@savepix \global\ymaxpix=\y@savepix
\fi}
\def\flushmove {\p@osinittrue
\ifm@pending
\writetx {\the\x@savepix\space \the\y@savepix\space mv}%
\m@pendingfalse
\p@athfalse
\fi}
\def\flushbs {\loop
\ifnum \d@bs>0
\writetx {bs}%
\global\advance \d@bs by -1
\repeat}
\def\h@move #1#2 #3)#4{\move (#2 #3)%
\h@text {#4}}
\def\h@text #1{\pixtodim \x@pix \t@xpos
\pixtodim \y@pix \t@ypos
\vbox to 0pt{\normalbaselines
\t@stuff
\kern -\t@ypos
\hbox to 0pt{\l@stuff
\kern \t@xpos
\hbox {#1}%
\kern -\t@xpos
\r@stuff}%
\kern \t@ypos
\b@stuff\relax}}
\def\r@move td:#1 #2#3 #4)#5{\move (#3 #4)%
\r@text td:#1 {#5}}
\def\r@text td:#1 #2{\pixtodim \x@pix \t@xpos
\pixtodim \y@pix \t@ypos
\vbox to 0pt{\kern -\t@ypos
\hbox to 0pt{\kern \t@xpos
\rottxt{#1}{#2}%
\hss}%
\vss}}
\def\rottxt #1#2{\rotsclTeX{#1}{1}{1}{\z@sb{#2}}}%
\def\z@sb #1{\vbox to 0pt{\normalbaselines
\t@stuff
\hbox to 0pt{\l@stuff
\hbox {#1}%
\r@stuff}%
\b@stuff}}
\def\t@exdrawdef {\sppix 300/in 
\drawdim in   
\edef\u@nitsc {1}%
\setsegscale 1    
\arrowheadsize l:0.16 w:0.08
\arrowheadtype t:T
\textref h:L v:B }
\def\writeps #1{\flushbs
\flushmove
\p@athtrue
\writetx {#1}}
\def\writetx #1{\ift@extonly
\t@extonlyfalse
\t@dropen
\fi
\w@rps {#1}}
\def\w@rps #1{\immediate\write\drawfile {#1}}
\def\t@dropen {%
\global\advance \t@xdnum by 1
\ifnum \t@xdnum<10
\xdef\p@sfile {\jobname.ps\the\t@xdnum}
\else
\xdef\p@sfile {\jobname.p\the\t@xdnum}
\fi
\immediate\openout\drawfile=\p@sfile
\w@rps {\p@b PS-Adobe-3.0 EPSF-3.0}%
\w@rps {\p@p BoundingBox: (atend)}%
\w@rps {\p@p Title: TeXdraw drawing: \p@sfile}%
\w@rps {\p@p Pages: 1 1}%
\w@rps {\p@p Creator: TeXdraw V1R3}%
\w@rps {\p@p CreationDate: \the\year/\the\month/\the\day}%
\w@rps {\p@p DocumentSuppliedResources: ProcSet TeXDraw 2.2 2}%
\w@rps {\p@p DocumentData: Clean7Bit}%
\w@rps {\p@p EndComments}%
\w@rps {\p@p BeginDefaults}%
\w@rps {\p@p PageNeededResources: ProcSet TeXDraw 2.2 2}%
\w@rps {\p@p EndDefaults}%
\w@rps {\p@p BeginProlog}%
\w@rps {\p@p BeginResource: ProcSet TeXDraw 2.2 2 14696 10668}%
\w@rps {\p@p VMlocation: local}%
\w@rps {\p@p VMusage: 14696 10668}%
\w@rps {/setglobal where}%
\w@rps {{pop currentglobal false setglobal} if}%
\w@rps {/setpacking where}%
\w@rps {{pop currentpacking false setpacking} if}%
\w@rps {29 dict dup begin}%
\w@rps {62 dict dup begin}%
\w@rps {/rad 0 def /radx 0 def /rady 0 def /svm matrix def}%
\w@rps {/hhwid 0 def /hlen 0 def /ah 0 def /tipy 0 def}%
\w@rps {/tipx 0 def /taily 0 def /tailx 0 def /dx 0 def}%
\w@rps {/dy 0 def /alen 0 def /blen 0 def}%
\w@rps {/i 0 def /y1 0 def /x1 0 def /y0 0 def /x0 0 def}%
\w@rps {/movetoNeeded 0 def}%
\w@rps {/y3 0 def /x3 0 def /y2 0 def /x2 0 def}%
\w@rps {/p1y 0 def /p1x 0 def /p2y 0 def /p2x 0 def}%
\w@rps {/p0y 0 def /p0x 0 def /p3y 0 def /p3x 0 def}%
\w@rps {/n 0 def /y 0 def /x 0 def}%
\w@rps {/anglefactor 0 def /elemlength 0 def /excursion 0 def}%
\w@rps {/endy 0 def /endx 0 def /beginy 0 def /beginx 0 def}%
\w@rps {/centery 0 def /centerx 0 def /startangle 0 def }%
\w@rps {/startradius 0 def /endradius 0 def /elemcount 0 def}%
\w@rps {/smallincrement 0 def /angleincrement 0 def /radiusincrement 0 def}%
\w@rps {/ifleft false def /ifright false def /iffill false def}%
\w@rps {/freq 1 def /angle 0 def /yrad 0 def /xrad 0 def /y 0 def /x 0 def}%
\w@rps {/saved 0 def}%
\w@rps {end}%
\w@rps {/dbdef {1 index exch 0 put 0 begin bind end def}}%
\w@rps {dup 3 4 index put dup 5 4 index put bind def pop}%
\w@rps {/bdef {bind def} bind def}%
\w@rps {/mv {stroke moveto} bdef}%
\w@rps {/lv {lineto} bdef}%
\w@rps {/st {currentpoint stroke moveto} bdef}%
\w@rps {/sl {st setlinewidth} bdef}%
\w@rps {/sd {st 0 setdash} bdef}%
\w@rps {/sg {st setgray} bdef}%
\w@rps {/bs {gsave} bdef /es {stroke grestore} bdef}%
\w@rps {/cv {curveto} bdef}%
\w@rps {/cr \l@br 0 begin}%
\w@rps {gsave /rad exch def currentpoint newpath rad 0 360 arc}%
\w@rps {stroke grestore end\r@br\space 0 dbdef}%
\w@rps {/fc \l@br 0 begin}%
\w@rps {gsave /rad exch def setgray currentpoint newpath}%
\w@rps {rad 0 360 arc fill grestore end\r@br\space 0 dbdef}%
\w@rps {/ar {gsave currentpoint newpath 5 2 roll arc stroke grestore} bdef}%
\w@rps {/el \l@br 0 begin gsave /rady exch def /radx exch def}%
\w@rps {svm currentmatrix currentpoint translate}%
\w@rps {radx rady scale newpath 0 0 1 0 360 arc}%
\w@rps {setmatrix stroke grestore end\r@br\space 0 dbdef}%
\w@rps {/fl \l@br gsave closepath setgray fill grestore}%
\w@rps {currentpoint newpath moveto\r@br\space bdef}%
\w@rps {/fp \l@br gsave closepath setgray fill grestore}%
\w@rps {currentpoint stroke moveto\r@br\space bdef}%
\w@rps {/av \l@br 0 begin /hhwid exch 2 div def /hlen exch def}%
\w@rps {/ah exch def /tipy exch def /tipx exch def}%
\w@rps {currentpoint /taily exch def /tailx exch def}%
\w@rps {/dx tipx tailx sub def /dy tipy taily sub def}%
\w@rps {/alen dx dx mul dy dy mul add sqrt def}%
\w@rps {/blen alen hlen sub def}%
\w@rps {gsave tailx taily translate dy dx atan rotate}%
\w@rps {(V) ah ne {blen 0 gt {blen 0 lineto} if} {alen 0 lineto} ifelse}%
\w@rps {stroke blen hhwid neg moveto alen 0 lineto blen hhwid lineto}%
\w@rps {(T) ah eq {closepath} if}%
\w@rps {(W) ah eq {gsave 1 setgray fill grestore closepath} if}%
\w@rps {(F) ah eq {fill} {stroke} ifelse}%
\w@rps {grestore tipx tipy moveto end\r@br\space 0 dbdef}%
\w@rps {/setupcurvy \l@br 0 begin}%
\w@rps {dup 0 eq {1 add} if /anglefactor exch def}%
\w@rps {abs dup 0 eq {1 add} if /elemlength exch def /excursion exch def}%
\w@rps {/endy exch def /endx exch def}%
\w@rps {/beginy exch def /beginx exch def}%
\w@rps {/centery exch def /centerx exch def}%
\w@rps {cleartomark}%
\w@rps {/startangle beginy centery sub beginx centerx sub atan def}%
\w@rps {/startradius beginy centery sub dup mul }%
\w@rps {beginx centerx sub dup mul add sqrt def}%
\w@rps {/endradius endy centery sub dup mul }%
\w@rps {endx centerx sub dup mul add sqrt def}%
\w@rps {endradius startradius sub }%
\w@rps {endy centery sub endx centerx sub atan }%
\w@rps {startangle 2 copy le {exch 360 add exch} if sub dup}%
\w@rps {elemlength startradius endradius add atan dup add}%
\w@rps {div round abs cvi dup 0 eq {1 add} if}%
\w@rps {dup /elemcount exch def }%
\w@rps {div dup anglefactor div dup /smallincrement exch def}%
\w@rps {sub /angleincrement exch def}%
\w@rps {elemcount div /radiusincrement exch def}%
\w@rps {gsave newpath}%
\w@rps {startangle dup cos startradius mul }%
\w@rps {centerx add exch }%
\w@rps {sin startradius mul centery add moveto}%
\w@rps {end \r@br 0 dbdef}%
\w@rps {/curvyphoton \l@br 0 begin}%
\w@rps {setupcurvy}%
\w@rps {elemcount \l@br /startangle startangle smallincrement add def}%
\w@rps {/startradius startradius excursion add def}%
\w@rps {startangle dup cos startradius mul }%
\w@rps {centerx add exch }%
\w@rps {sin startradius mul centery add}%
\w@rps {/excursion excursion neg def}%
\w@rps {/startangle startangle angleincrement add }%
\w@rps {smallincrement sub def}%
\w@rps {/startradius startradius radiusincrement add def}%
\w@rps {startangle dup cos startradius mul }%
\w@rps {centerx add exch }%
\w@rps {sin startradius mul centery add}%
\w@rps {/startradius startradius excursion add def}%
\w@rps {/startangle startangle smallincrement add def}%
\w@rps {startangle dup cos startradius mul }%
\w@rps {centerx add exch }%
\w@rps {sin startradius mul centery add curveto\r@br repeat}%
\w@rps {stroke grestore end}%
\w@rps {\r@br 0 dbdef}%
\w@rps {/curvygluon \l@br 0 begin}%
\w@rps {setupcurvy /radiusincrement radiusincrement 2 div def}%
\w@rps {elemcount \l@br startangle angleincrement add dup}%
\w@rps {cos startradius mul centerx add exch}%
\w@rps {sin startradius mul centery add}%
\w@rps {/startradius startradius radiusincrement add}%
\w@rps {excursion sub def}%
\w@rps {startangle angleincrement add dup}%
\w@rps {cos startradius mul centerx add exch}%
\w@rps {sin startradius mul centery add}%
\w@rps{startangle angleincrement smallincrement add 2 div add dup}%
\w@rps {cos startradius mul centerx add exch}%
\w@rps {sin startradius mul centery add}%
\w@rps {curveto}%
\w@rps {/startangle startangle angleincrement smallincrement add add def}%
\w@rps {startangle angleincrement sub dup}%
\w@rps {cos startradius mul centerx add exch}%
\w@rps {sin startradius mul centery add}%
\w@rps {/startradius startradius radiusincrement add}%
\w@rps {excursion add def}%
\w@rps {startangle angleincrement sub dup}%
\w@rps {cos startradius mul centerx add exch}%
\w@rps {sin startradius mul centery add}%
\w@rps {startangle dup}%
\w@rps {cos startradius mul centerx add exch}%
\w@rps {sin startradius mul centery add}%
\w@rps {curveto\r@br repeat}%
\w@rps {stroke grestore end}%
\w@rps {\r@br 0 dbdef}%
\w@rps {/blob \l@br}%
\w@rps {0 begin st gsave}%
\w@rps {dup type dup}%
\w@rps {/stringtype eq}%
\w@rps {\l@br pop 0 get }%
\w@rps {dup (B) 0 get eq dup 2 index}%
\w@rps {(L) 0 get eq or /ifleft exch def}%
\w@rps {exch (R) 0 get eq or /ifright exch def}%
\w@rps {/iffill false def \r@br}%
\w@rps {\l@br /ifleft false def}%
\w@rps {/ifright false def}%
\w@rps {/booleantype eq }%
\w@rps {{/iffill exch def}}%
\w@rps {{setgray /iffill true def} ifelse \r@br}%
\w@rps {ifelse}%
\w@rps {/freq exch def}%
\w@rps {/angle exch def}%
\w@rps {/yrad  exch def}%
\w@rps {/xrad  exch def}%
\w@rps {/y exch def}%
\w@rps {/x exch def}%
\w@rps {newpath}%
\w@rps {svm currentmatrix pop}%
\w@rps {x y translate   }%
\w@rps {angle rotate}%
\w@rps {xrad yrad scale}%
\w@rps {0 0 1 0 360 arc}%
\w@rps {gsave 1 setgray fill grestore}%
\w@rps {gsave svm setmatrix stroke grestore}%
\w@rps {gsave iffill {fill} if grestore}%
\w@rps {clip newpath}%
\w@rps {gsave }%
\w@rps {ifleft  \l@br -3 freq 3 { -1 moveto 2 2 rlineto} for}%
\w@rps {svm setmatrix stroke\r@br if }%
\w@rps {grestore}%
\w@rps {ifright \l@br 3 freq neg -3 { -1 moveto -2 2 rlineto} for}%
\w@rps {svm setmatrix stroke\r@br if}%
\w@rps {grestore end}%
\w@rps {\r@br 0 dbdef}%
\w@rps {/BSpl \l@br}%
\w@rps {0 begin}%
\w@rps {storexyn}%
\w@rps {currentpoint newpath moveto}%
\w@rps {n 1 gt \l@br}%
\w@rps {0 0 0 0 0 0 1 1 true subspline}%
\w@rps {n 2 gt \l@br}%
\w@rps {0 0 0 0 1 1 2 2 false subspline}%
\w@rps {1 1 n 3 sub \l@br}%
\w@rps {/i exch def}%
\w@rps {i 1 sub dup i dup i 1 add dup i 2 add dup false subspline}%
\w@rps {\r@br for}%
\w@rps {n 3 sub dup n 2 sub dup n 1 sub dup 2 copy false subspline}%
\w@rps {\r@br if}%
\w@rps {n 2 sub dup n 1 sub dup 2 copy 2 copy false subspline}%
\w@rps {\r@br if}%
\w@rps {end}%
\w@rps {\r@br 0 dbdef}%
\w@rps {/midpoint \l@br}%
\w@rps {0 begin}%
\w@rps {/y1 exch def}%
\w@rps {/x1 exch def}%
\w@rps {/y0 exch def}%
\w@rps {/x0 exch def}%
\w@rps {x0 x1 add 2 div}%
\w@rps {y0 y1 add 2 div}%
\w@rps {end}%
\w@rps {\r@br 0 dbdef}%
\w@rps {/thirdpoint \l@br}%
\w@rps {0 begin}%
\w@rps {/y1 exch def}%
\w@rps {/x1 exch def}%
\w@rps {/y0 exch def}%
\w@rps {/x0 exch def}%
\w@rps {x0 2 mul x1 add 3 div}%
\w@rps {y0 2 mul y1 add 3 div}%
\w@rps {end}%
\w@rps {\r@br 0 dbdef}%
\w@rps {/subspline \l@br}%
\w@rps {0 begin}%
\w@rps {/movetoNeeded exch def}%
\w@rps {y exch get /y3 exch def}%
\w@rps {x exch get /x3 exch def}%
\w@rps {y exch get /y2 exch def}%
\w@rps {x exch get /x2 exch def}%
\w@rps {y exch get /y1 exch def}%
\w@rps {x exch get /x1 exch def}%
\w@rps {y exch get /y0 exch def}%
\w@rps {x exch get /x0 exch def}%
\w@rps {x1 y1 x2 y2 thirdpoint}%
\w@rps {/p1y exch def}%
\w@rps {/p1x exch def}%
\w@rps {x2 y2 x1 y1 thirdpoint}%
\w@rps {/p2y exch def}%
\w@rps {/p2x exch def}%
\w@rps {x1 y1 x0 y0 thirdpoint}%
\w@rps {p1x p1y midpoint}%
\w@rps {/p0y exch def}%
\w@rps {/p0x exch def}%
\w@rps {x2 y2 x3 y3 thirdpoint}%
\w@rps {p2x p2y midpoint}%
\w@rps {/p3y exch def}%
\w@rps {/p3x exch def}%
\w@rps {movetoNeeded \l@br p0x p0y moveto \r@br if}%
\w@rps {p1x p1y p2x p2y p3x p3y curveto}%
\w@rps {end}%
\w@rps {\r@br 0 dbdef}%
\w@rps {/storexyn \l@br}%
\w@rps {0 begin}%
\w@rps {/n exch def}%
\w@rps {/y n array def}%
\w@rps {/x n array def}%
\w@rps {n 1 sub -1 0 \l@br}%
\w@rps {/i exch def}%
\w@rps {y i 3 2 roll put}%
\w@rps {x i 3 2 roll put}%
\w@rps {\r@br for end}%
\w@rps {\r@br 0 dbdef}%
\w@rps {/bop \l@br save 0 begin /saved exch def end}%
\w@rps {scale setlinecap setlinejoin setlinewidth setdash moveto}%
\w@rps {\r@br 1 dbdef}%
\w@rps {/eop {stroke 0 /saved get restore showpage} 1 dbdef}%
\w@rps {end /defineresource where}%
\w@rps {{pop mark exch /TeXDraw exch /ProcSet defineresource cleartomark}}%
\w@rps {{/TeXDraw exch readonly def} ifelse}%
\w@rps {/setpacking where {pop setpacking} if}%
\w@rps {/setglobal where {pop setglobal} if}%
\w@rps {\p@p EndResource}%
\w@rps {\p@p EndProlog}%
\w@rps {\p@p Page: 1 1}%
\w@rps {\p@p PageBoundingBox: (atend)}%
\w@rps {\p@p BeginPageSetup}%
\w@rps {/TeXDraw /findresource where}%
\w@rps {{pop /ProcSet findresource}}%
\w@rps {{load} ifelse}%
\w@rps {begin}%
\w@rps {0 0 [] 0 3 1 1 \p@sfactor\space \p@sfactor\space bop}%
\w@rps {\p@p EndPageSetup}%
}
\def\t@drclose {%
\pixtobp \xminpix \l@lxbp  \pixtobp \yminpix \l@lybp
\pixtobp \xmaxpix \u@rxbp  \pixtobp \ymaxpix \u@rybp
\w@rps {\p@p PageTrailer}%
\w@rps {\p@p PageBoundingBox: \the\l@lxbp\space \the\l@lybp\space
\the\u@rxbp\space \the\u@rybp}%
\w@rps {eop end}%
\w@rps {\p@p Trailer}%
\w@rps {\p@p BoundingBox: \the\l@lxbp\space \the\l@lybp\space
\the\u@rxbp\space \the\u@rybp}%
\w@rps {\p@p EOF}%
\closeout\drawfile}
\catcode`\@=\catamp
\def\dvialwsetup{
\def\includepsfile##1##2##3##4##5{\special{Insert ##1\space%
}}%
\def\rotsclTeX##1##2##3##4{\special{Insert /dev/null do %
3 index exch translate cleartomark %
matrix currentmatrix aload pop %
7 6 roll restore matrix astore %
matrix currentmatrix exch setmatrix %
0 0 moveto setmatrix %
gsave currentpoint 2 copy translate ##1 rotate %
##2 ##3 scale neg exch neg exch translate %
save}%
##4%
\special{Insert /dev/null do cleartomark restore %
currentpoint grestore moveto save}}%
}
\def\dvipssetup{
\def\includepsfile##1##2##3##4##5{\vbox to 0pt{%
\vskip##5%
\includegraphics{##1}%
\vss}}
\def\rotsclTeX##1##2##3##4{%
##4
}}
\dvipssetup


\chardef\catamp=\the\catcode`\@
\catcode`\@=11
\def\realadd #1#2#3{\dimen0=#1pt
\dimen2=#2pt
\advance \dimen0 by \dimen2
\edef #3{\expandafter\c@lean\the\dimen0}}
\def\realdiv #1#2#3{\dimen0=#1pt
\t@counta=\dimen0
\dimen0=#2pt
\t@countb=\dimen0
\intdiv \t@counta \t@countb #3}
\def\lenhyp #1#2#3{\t@counta=#1%
\multiply \t@counta by \t@counta
\t@countb=#2%
\multiply \t@countb by \t@countb
\advance \t@counta by \t@countb
\sqrtnum \t@counta #3}
\let\bk=\t@counta
\let\bn=\t@countb
\let\xval=\t@countc
\def\sqrtnum #1#2{\xval=#1%
\bk=\xval
\loop
\bn=\xval
\divide \bn by \bk
\advance \bn by \bk
\advance \bn by 1
\divide \bn by 2
\ifnum \bn < \bk
\bk=\bn
\repeat
#2=\bn}
\def\currentpos #1#2{\t@pixa=\x@pix
\advance \t@pixa by -\x@segoffpix
\pixtocoord \t@pixa #1
\t@pixa=\y@pix
\advance \t@pixa by -\y@segoffpix
\pixtocoord \t@pixa #2}
\def\vectlen (#1 #2)(#3 #4)#5{\getpos (#1 #2)\x@arga\y@arga
\getpos (#3 #4)\x@argb\y@argb
\coordtopix \x@arga \t@pixa
\coordtopix \x@argb \t@pixb
\advance \t@pixb by -\t@pixa
\coordtopix \y@arga \t@pixc
\coordtopix \y@argb \t@pixd
\advance \t@pixd by -\t@pixc
\lenhyp \t@pixb \t@pixd \t@pixc
\pixtocoord \t@pixc #5}
\def\cossin (#1 #2)(#3 #4)#5#6{\getpos (#1 #2)\x@arga\y@arga
\getpos (#3 #4)\x@argb\y@argb
\coordtopix \x@arga \t@pixa
\coordtopix \x@argb \t@pixb
\advance \t@pixb by -\t@pixa
\coordtopix \y@arga \t@pixc
\coordtopix \y@argb \t@pixd
\advance \t@pixd by -\t@pixc
\lenhyp \t@pixb \t@pixd \t@pixc
\intdiv \t@pixb\t@pixc #5%
\intdiv \t@pixd\t@pixc #6}
\catcode`\@=\catamp

\expandafter\chardef\csname m@catamp\endcsname=\the\catcode`\@
\catcode`\@=11
\def\straightness #1 {\realdiv 1{#1}\m@factor}

\def\gvec (#1 #2) {\currentpos \m@xpos\m@ypos
\realadd {#1}\m@xpos\m@
\realdiv \m@2\m@xav
\realadd {#2}\m@ypos\m@
\realdiv \m@2\m@yav
\edef\m@{(\m@xpos\space\m@ypos)}
\expandafter\cossin \m@(#1 #2)\m@cos\m@sin
\realdiv \m@sin{-\m@factor}\m@
\realadd \m@\m@xav\m@xcntr
\realdiv \m@cos\m@factor\m@
\realadd \m@\m@yav\m@ycntr
\edef\m@{around (\m@xcntr\space\m@ycntr) %
from (\m@xpos\space\m@ypos) to (#1 #2) }
\expandafter\drawcurvygluon \m@}

\def\pvec (#1 #2) {\currentpos \m@xpos\m@ypos
\realadd {#1}\m@xpos\m@
\realdiv \m@2\m@xav
\realadd {#2}\m@ypos\m@
\realdiv \m@2\m@yav
\edef\m@{(\m@xpos\space\m@ypos)}
\expandafter\cossin \m@(#1 #2)\m@cos\m@sin
\realdiv \m@sin{-\m@factor}\m@
\realadd \m@\m@xav\m@xcntr
\realdiv \m@cos\m@factor\m@
\realadd \m@\m@yav\m@ycntr
\edef\m@{around (\m@xcntr\space\m@ycntr) %
from (\m@xpos\space\m@ypos) to (#1 #2) }
\expandafter\drawcurvyphoton \m@}

\straightness{10}
\catcode`\@=\m@catamp
\straightness 100 


\newpage
\thispagestyle{empty}

\newcommand{\grSE}{
\begin{texdraw} \drawdim cm \setunitscale 0.29 \curvylength {15}
\curvyheight {10} \arrowheadtype t:F \arrowheadsize l:0.6 w:0.3
\move (-5 4) \lvec (-5 -4) \move (-5 2) \avec (-5 2.5) \move (-5 -2.5)
\avec (-5 -2) \move (5 4) \lvec (5 -4) \move (5 2) \avec (5 2.5)
\move (5 -2.5) \avec (5 -2) \move (-5 0) \pvec (0 0) \pvec (5 0)
\grayblob {0.8} \drawblob xsize:2.5 ysize:2.5 at (0 0) \move (0 0)
\lcir r:2.5 
\htext (-7 1.5) {$f$} \htext (-7 -3) {$i$} \htext (6.5 1.5) {$f'$}
\htext (6.5 -3) {$i'$} \htext (-4 -2) {$k$} \htext (-1 -6) {(a)}
\end{texdraw}}

\newcommand{\grSEB}{
\begin{texdraw} \drawdim cm \setunitscale 0.25 \curvylength {15}
\curvyheight {10} \arrowheadtype t:F \arrowheadsize l:0.6 w:0.5
\drawcurvyphoton around (0 0) from (-2.3 0) to (2.3 0)
\pvec (6 0)
\drawcurvyphoton around (0 0) from (2.3 0) to (-2.3 0)
\pvec (-6 0)
\htext (-4 -2) {$k$} 
\htext (-1 -6) {(b)}
\end{texdraw}}

\newcommand{\grSEC}{
\begin{texdraw} \drawdim cm \setunitscale 0.25 \curvylength {15}
\curvyheight {10} \arrowheadtype t:F \arrowheadsize l:0.6 w:0.3
\move (-6 -1) \pvec (0 -1) \move (0 2)
\drawcurvyphoton around (-0.25 1.25) from (1.25 2.25) to (1.25 2.25)
\move (0 -1) \pvec (6 -1)
\htext (0 -3) {$k$} 
\htext (-1 -6) {(c)}
\end{texdraw}}

\newcommand{\grSED}{
\begin{texdraw} \drawdim cm \setunitscale 0.25 \curvylength {15}
\curvyheight {10} \arrowheadtype t:F \arrowheadsize l:0.6 w:0.5
\move (-6 0) \pvec (-2.3 0) \move (0 0) \lpatt(.7 .3) \lcir r:2.3 \lpatt()
\move (2.3 0) \pvec (6 0) \move (0 2.3) \avec (0.5 2.3) \move (0 -2.3)
\avec (-0.5 -2.3)
\htext (-4 -2) {$k$} \htext (-1 -6) {(d)}
\end{texdraw}}


\vbox to \vsize{
\tabskip=0cmplus1fil
\halign to \hsize{&\hfil#\hfil\cr
\noalign{\vskip 0cmplus1fil}
                &\grSE          &\cr
\noalign{\vskip 3cm}
        \grSEB  &\grSEC         &\grSED\cr
\noalign{\vskip 0cmplus1fil}
\noalign{\vskip 0cmplus1fil}
                &Figure~1       &\cr
}}

\newpage
\thispagestyle{empty}

\newcommand{\grVertexFirstone}{\begin{texdraw} \drawdim cm \setunitscale 0.25
\curvylength {15} \curvyheight {10} \arrowheadtype t:F 
\arrowheadsize l:0.6 w:0.3 \move (-5 4) \lvec (-5 -4) \move (5 4) \lvec (5 -4)
\drawcurvyphoton around (-5 0) from (-5 2.5) to (-5 -2.5)
\move (-5 0) \pvec (0 0) \pvec (5 0)
\htext (0 -2) {$k$} \htext (-9 0) {$p$} \htext (9 0) {+} 
\htext (9 -8) {(a)} 
\end{texdraw}}

\newcommand{\grVertexFirsttwo}{\begin{texdraw} \drawdim cm \setunitscale 0.25
\curvylength {15}
\curvyheight {10} \arrowheadtype t:F \arrowheadsize l:0.6 w:0.3
\move (-5 4) \lvec (-5 -4) \move (5 4) \lvec (5 -4)
\drawcurvyphoton around (-5 0.75) from (-5 2.75) to (-5 -1.25)
\move (-5 -2.5) \pvec (0 -2.5) \pvec (5 -2.5)
\htext (-8 -2) {$p$} \htext (0 -4.5) {$k$}
\htext (10 0) {$\Longrightarrow$} 
\htext (10 -8) {}
\end{texdraw}}

\newcommand{\grPinchVertexFirst}{\begin{texdraw} \drawdim cm \setunitscale 0.25
\curvylength {15} \curvyheight {10} \arrowheadtype t:F
\arrowheadsize l:0.6 w:0.3
\move (-1 4) \lvec (-1 -4) \move (9 4) \lvec (9 -4)
\drawcurvyphoton around (-3.3 0) from (-3.3 -2.1) to (-3.3 -2.1)
\grayblob {0.0}
\drawblob xsize:0.2 ysize:0.2 at (-1 0)
\move (-1 0) \pvec (4 0) \pvec (9 0)
\htext (4 -2) {$k$} \htext (-3 -4) {$p$} \htext (4 -8) {(b)} \htext (-7 0) {}
\end{texdraw}}



\newcommand{\grVertexSecond}{\begin{texdraw} \drawdim cm \setunitscale 0.25 
\curvylength {15}
\curvyheight {10} \arrowheadtype t:F \arrowheadsize l:0.6 w:0.3
\move (-5 4) \lvec (-5 -4) \move (5 4) \lvec (5 -4)
\move (-5 2.3) \pvec (-1.6 0) \pvec (2.5 0) \pvec (5 0) \move (-1.6 0)
\pvec (-5 -2.3)
\htext (3 -2) {$k$} \htext (-3 2) {$p+k$} \htext (-3 -3) {$p$} 
\htext (0 -8) {(a)} \htext (20 0) {$\Longrightarrow$} 
\end{texdraw}}


\newcommand{\grPinchVertexSecond}{\begin{texdraw} \drawdim cm 
\setunitscale 0.25 \curvylength {15} \curvyheight {10} \arrowheadtype t:F 
\arrowheadsize l:0.6 w:0.3
\move (-5 4) \lvec (-5 -4) \move (5 4) \lvec (5 -4)
\drawcurvyphoton around (-2.5 1.25) from (-5 0) to (0 0)
\drawcurvyphoton around (-2.5 -1.25) from (0 0) to (-5 0)
\grayblob {0.0} \drawblob xsize:0.2 ysize:0.2 at (-5 0)
\move (0 0) \pvec (5 0) 
\htext (3 -2) {$k$} \htext (-2.5 -3) {$p$} \htext (-3.5 2.5) {$p+k$}
\htext (0 -8) {(b)}
\end{texdraw}}


\newcommand{\grBoxone}{\begin{texdraw} \drawdim cm \setunitscale 0.25 
\curvylength {15}
\curvyheight {10} \arrowheadtype t:F \arrowheadsize l:0.6 w:0.3
\move (-5 4) \lvec (-5 -4) \move (5 4) \lvec (5 -4)
\move (-5 2) \pvec (0 2) \pvec (5 2) \move (-5 -2) \pvec (0 -2) \pvec (5 -2)
\htext (-0.5 -4) {$p$} \htext (-1.5 3) {$p+k$} \htext (10 0) {+}
\htext (10 -8) {(a)}
\end{texdraw}}

\newcommand{\grBoxtwo}{\begin{texdraw} \drawdim cm \setunitscale 0.25 
\curvylength {15}
\curvyheight {10} \arrowheadtype t:F \arrowheadsize l:0.6 w:0.3
\move (-5 4) \lvec (-5 -4) \move (5 4) \lvec (5 -4) \move (-5 2)
\pvec (0 0) \pvec (5 -2) \move (5 2) \pvec (0 0) \pvec (-5 -2)
\htext (-2 -2.5) {$p$} \htext (-3 2) {$p+k$} 
\htext (13 0) {$\Longrightarrow$}
\htext (0 -8) {}
\end{texdraw}}

\newcommand{\grBoxP}{ \begin{texdraw} \drawdim cm \setunitscale 0.25 
\curvylength {15}
\curvyheight {10} \arrowheadtype t:F \arrowheadsize l:0.6 w:0.3
\move (-5 4) \lvec (-5 -4) \move (5 4) \lvec (5 -4)
\drawcurvyphoton around (0 -3) from (5 0) to (-5 0)
\drawcurvyphoton around (0 3) from (-5 0) to (5 0)
\grayblob {0.0} \drawblob xsize:0.2 ysize:0.2 at (-5 0)
\grayblob {0.0} \drawblob xsize:0.2 ysize:0.2 at (5 0)
\htext (0 -4.5) {$p$} \htext (-1.5 4) {$p+k$} 
\htext (0 -8) {(b)}
\end{texdraw}}


\vbox to \vsize{
\tabskip=0cmplus1fil
\halign to \hsize{#&#&#\cr
\noalign{\vskip 0cmplus1fil}
\grVertexFirstone & \hspace{-1.5cm}\grVertexFirsttwo    &\hspace{1cm}
                                                 \grPinchVertexFirst \cr
\noalign{\vskip 0cmplus1fil}
                        &$\mbox{Figure~2}$              &\cr
\noalign{\vskip 0cmplus1fil}
        \qquad\quad\grVertexSecond &            &\qquad\grPinchVertexSecond \cr
\noalign{\vskip 0cmplus1fil}
                        &$\mbox{Figure~3}$              &\cr
\noalign{\vskip 0cmplus1fil}
        \grBoxone       & \hspace{-1.5cm}\grBoxtwo &\quad\grBoxP \cr
\noalign{\vskip 0cmplus1fil}
                        &$\mbox{Figure~4}$              &\cr
\noalign{\vskip 0cmplus1fil}
}}

\newpage
\thispagestyle{empty}

\newcommand{\CGglghVertex}{
\begin{texdraw} \drawdim cm \setunitscale 0.35 \curvylength {15}
\curvyheight {10} \arrowheadtype t:F \arrowheadsize l:0.8 w:0.4
\move (-6.1 -3) \lpatt(.7 .3) \lvec(0 0) \move(6.1 -3) \lvec(0 0) \lpatt()
\pvec(0 6) \move(-3.2 -1.6) \avec(-2.6 -1.3) \move(3.6 -1.8) \avec(4 -2)
\arrowheadsize l:0.9 w:0.45
\move(-4.14 -2.8) \avec(-1.5 -1.5)
\move(4.14 -2.8) \avec(1.5 -1.5)
\move(-2.2 -3.3) \htext{$q$} \move(1.3 -3.3)\htext{$p$}
\move(-2 4.8) \htext{$\mu$} \move(1 5.1) \htext{$b$}
\move(-6.3 -2.2) \htext{$c$} \move(5.4 -2.2) \htext{$a$}
\move(-0.8 3.8) \avec(-0.8 0.9) \move(-2.2 2.2)\htext{$k$}
\move(4 1)\htext{$\Gamma_{\mu}^{abc}(p,k,q)=g f^{abc}[p_{\mu}-p_0 n_{\mu}]$
        ~~~~~~~(a)}
\end{texdraw}}

\newcommand{\TAGglghVertex}{
\begin{texdraw} \drawdim cm \setunitscale 0.35 \curvylength {15}
\curvyheight {10} \arrowheadtype t:F \arrowheadsize l:0.8 w:0.4
\move (-6.1 -3) \lpatt(.7 .3) \lvec(0 0) \move(6.1 -3) \lvec(0 0) \lpatt()
\pvec(0 6) \move(-3.2 -1.6) \avec(-2.6 -1.3) \move(3.6 -1.8) \avec(4 -2)
\arrowheadsize l:0.9 w:0.45
\move(-4.14 -2.8) \avec(-1.5 -1.5)
\move(4.14 -2.8) \avec(1.5 -1.5)
\move(-2.2 -3.3) \htext{$q$} \move(1.3 -3.3)\htext{$p$}
\move(-2 4.8) \htext{$\mu$} \move(1 5.1) \htext{$b$}
\move(-6.3 -2.2) \htext{$c$} \move(5.4 -2.2) \htext{$a$}
\move(-0.8 3.8) \avec(-0.8 0.9) \move(-2.2 2.2)\htext{$k$}
\move(4 1) \htext{$\Gamma_{\mu}^{abc}(p,k,q)=-ig f^{abc}[n_{\mu}]$
        ~~~~~~~~~~~~~~~(b)}
\end{texdraw}}

\vbox to \vsize{
\tabskip=0cmplus1fil
\halign to \hsize{&\hfil#\hfil\cr
\CGglghVertex  &        &\cr
\noalign{\vskip 0cmplus1fil}
        \TAGglghVertex &        &\cr
\noalign{\vskip 0cmplus1fil}
                &Figure~5       &\cr
}}

\end{document}